\documentclass[PAPER, cernpreprint, atlasdraft=false, texlive=2023, UKenglish, texmf, orcidlogo]{atlasdoc}
 
\usepackage{atlaspackage}
\usepackage{needspace}
\usepackage{atlasbiblatex}
 
\usepackage{atlasphysics}

\graphicspath{{logos/}{figures/}}
 
\usepackage{ANA-TOPQ-2023-21-PAPER-defs}

\AtlasTitle{Measurement of the $t\bar{t}$ cross section and its ratio to the $Z$ production cross section using $pp$ collisions at $\sqrt{s} = 13.6$~TeV with the ATLAS detector}
 
\AtlasVersion{1.5}
 
\AtlasAbstract{
The inclusive top-quark-pair production cross section $\sigma_{t\bar{t}}$ and
its ratio to the $Z$-boson production cross section have been measured in proton--proton collisions at $\sqrt{s} = 13.6$~TeV, using 29~fb${}^{-1}$ of data collected in 2022 with the ATLAS experiment at the Large Hadron Collider.
Using events with an opposite-charge electron-muon pair and $b$-tagged jets, and assuming Standard Model decays,
the top-quark-pair production cross section is measured to be $\sigma_{t\bar{t}} = 850 \pm 3\mathrm{(stat.)}\pm 18\mathrm{(syst.)}\pm 20\mathrm{(lumi.)}$~pb.
The ratio of the $t\bar{t}$ and the $Z$-boson production cross sections is also measured, where the $Z$-boson contribution is determined for inclusive $e^+e^-$ and $\mu^+\mu^-$ events in a fiducial phase space.
The relative uncertainty on the ratio is reduced compared to the $t\bar{t}$ cross section, thanks to the cancellation of several systematic uncertainties.
The result for the ratio, $R_{t\bar{t}/Z} = 1.145 \pm 0.003\mathrm{(stat.)}\pm 0.021\mathrm{(syst.)}\pm 0.002\mathrm{(lumi.)}$ is consistent with
the Standard Model prediction using the PDF4LHC21 PDF set.
}
 
\author{The ATLAS Collaboration}
 
\AtlasRefCode{TOPQ-2023-21}
 
\PreprintIdNumber{CERN-EP-2023-166}

\AtlasJournal{Phys.\ Lett.\ B.}
\AtlasJournalRef{\PLB 848 (2024) 138376}
\AtlasDOI{10.1016/j.physletb.2023.138376}

\hypersetup{pdftitle={ATLAS document},pdfauthor={The ATLAS Collaboration}}
 
\begin{document}
 
\maketitle
 
\tableofcontents

% The next lines are included from the .//Introduction.tex input file
\section{Introduction}
\label{sec:intro}
 
Measurements of the top-quark production cross section in hadron-hadron
collisions at the LHC provide essential
tests of Quantum Chromodynamics (QCD) and parton
distributions functions (PDFs).
This quantity has been measured by
ATLAS~\cite{TOPQ-2018-40-fix,TOPQ-2013-04,TOPQ-2011-02,TOPQ-2018-26}
and
CMS~\cite{CMS-TOP-16-023,CMS-TOP-13-004,CMS-TOP-12-006,CMS-TOP-11-003,CMS-TOP-15-003}
at energies between 5 and 13~\tev, and by
LHCb~\cite{LHCb:2015nta,LHCb:2018usb} at energies between 7 and
13~\tev, with percent-level precision.
Theory calculations with similar precision
are available
\cite{topxtheo1,topxtheo2,topxtheo3,topxtheo4,topxtheo5,topxtheo6}
and are found to be in good agreement with the measurements.
 
In the ratio of the \ttbar\ to the $Z$-boson production cross
section, several experimental uncertainties cancel, so
that it can be obtained very precisely. Since $\ttbar$ production proceeds
preferentially via a gluon-gluon initial state while $Z$-boson production is
dominated by $\qqbar$,
the ratio of these cross sections at a given centre-of-mass energy has a significant sensitivity to the gluon-to-quark
PDF ratio~\cite{Rojo:2015acz,Mangano:2012mh}.
ATLAS published such ratios at various centre-of-mass energies in Ref.~\cite{STDM-2016-02}.
Predictions for the $Z$-boson production cross section are available at
NNLO QCD plus next-to-leading-order (NLO) electroweak (EW)
accuracy~\cite{Anastasiou:2003ds,Catani:2007vq,Catani:2009sm,Gavin:2010az-fix,Li:2012wna}.
 
The analysis uses proton--proton collision data delivered by the LHC at CERN
in 2022 at a new centre-of-mass
energy of $\sqrt{s}=13.6$~\tev and recorded with the ATLAS detector, representing an integrated luminosity of 29~\ifb.
Due to the steep rise
of the PDFs towards lower Bjorken-$x$, a 12\% increase of the
$\ttbar$ cross section at 13.6~\tev is expected with respect
to 13~\tev, which should be visible already with the present precision on the detector calibration.
 
This letter reports the measurement of the \ttbar\ cross section in the full
phase-space and of the ratio of the $\ttbar$ and $Z$-boson cross sections.
The CMS experiment has published a measurement of the $\ttbar$ cross section at $\sqrt{s}=13.6\,\tev$~\cite{CMS:2023qyl} using an integrated luminosity of 1.21~\ifb, without including the ratio to the $Z$-boson cross section.

\FloatBarrier

% End of text imported from the .//Introduction.tex input file

% The next lines are included from the .//Strategy.tex input file
\section{Analysis strategy}
\label{sec:strategy}
 
Top-quark pair production is analysed in the channel in which one top-quark decays
into an electron, a neutrino and a $b$-quark, and the other top-quark into a
muon, a neutrino and a $b$-quark.
For $Z$-bosons, decays to a pair of electrons or muons are considered.
Three channels are thus defined using electrons and muons of opposite electric charge: $ee$, $\mu\mu$ and $e\mu$.
 
In the $e\mu$ channel, the presence of additional
$b$-tagged jets is required. The analysis uses the method already
employed in Ref.~\cite{TOPQ-2013-04,TOPQ-2018-17}, where the numbers of events with one and two $b$-tagged jets
are counted separately, allowing the $\ttbar$ cross section to be extracted
together with the efficiency to identify a $b$-tagged jet.
Opposite-sign electron-muon pairs are selected and the numbers of events that have one $b$-tagged jet ($N_1$) and two $b$-tagged jets ($N_2$) are counted.
The event yields are given by
 
\begin{eqnarray}
\label{eq:btagcount}
N_1 &=& L\sigma_{\ttbar}\epsilon_{e\mu}2\epsilon_b\left(1-C_b\epsilon_b\right) + N_1^{\text{bkg}},\\
N_2 &=& L\sigma_{\ttbar}\epsilon_{e\mu}C_b\epsilon_b^2 + N_2^{\text{bkg}},
\end{eqnarray}
 
where $L$ is the integrated luminosity, $\sigma_{\ttbar}$ is the
measured \ttbar cross section, $\epsilon_b$ is the combined probability to
reconstruct and $b$-tag a $b$-jet after the selection, and $\epsilon_{e\mu}$ is
the efficiency for a \ttbar event to pass the opposite-sign $e\mu$
selection.
The tagging correlation coefficient $C_b= \epsilon_{bb}/\epsilon_b^2$,
where  $\epsilon_{bb}$ represents the probability to reconstruct and tag
both $b$-jets, may deviate slightly from unity due to kinematic
correlations of the $b$-jets produced in a $\ttbar$ event and is estimated
from simulation.
The nominal value for $C_b$ is found to be 1.0068.
 
For the same-flavour channels ($ee$, $\mu\mu$), the reconstructed invariant mass of the lepton pair, $m_{\ell\ell}$, is required to be $66 < m_{\ell\ell} < 116$~\gev.
These channels are then used to extract the cross section for $Z$-boson production
 
\begin{eqnarray}
\label{eq:sigz}
N_{ee} &=& L\sigma_{Z}\epsilon_{ee} + N_{ee}^{\text{bkg}},\\
N_{\mu\mu} &=& L\sigma_{Z}\epsilon_{\mu\mu} + N_{\mu\mu}^{\text{bkg}},
\end{eqnarray}
 
where $N_{ee}$ ($N_{\mu\mu}$) is the selected number of events passing the $ee$ ($\mu\mu$) selection.
Parameters $\epsilon_{ee}$ and $\epsilon_{\mu\mu}$ are the efficiencies for a $Z$-boson event to pass the opposite-sign $ee$ and $\mu\mu$ selection, respectively.
 
The fiducial phase-space for $Z$-boson production is defined by the
transverse momentum of the leptons $\pt > 27~\gev$, the absolute value of the leptons' pseudorapidity $|\eta| < 2.5$ and the invariant mass of $66 < m_{\ell\ell} < 116~\gev$ for $\ell = e, \mu$.\footnote{\AtlasCoordFootnote}
Within this phase space, an almost pure $Z$-boson contribution is obtained, with about 99.5\% of events in the simulation originating from $Z$-boson production.

The total \ttbar production
cross section and the ratio of the cross sections, \ratio, are measured in
a common fit to the same- and different-flavour channels, using a binned profile-likelihood technique.
To understand the cancellation of uncertainties in the cross section ratio,
the fiducial $Z$-boson production cross section \xz\ is measured in a separate fit using same-flavour dilepton events only, within the same phase
space as used for the ratio measurement.
 
\FloatBarrier

% End of text imported from the .//Strategy.tex input file

% The next lines are included from the .//Detector.tex input file
\section{ATLAS detector}
\label{sec:ATLAS}
\needspace{7\baselineskip}
The ATLAS experiment~\cite{PERF-2007-01,GENR-2019-02} at the LHC is a multipurpose particle detector
with a forward--backward symmetric cylindrical geometry and a near \(4\pi\) coverage in
solid angle.
It consists of an inner tracking detector (ID) surrounded by a thin superconducting solenoid
providing a \SI{2}{\tesla} axial magnetic field, electromagnetic and hadron calorimeters, and a muon spectrometer (MS).
The inner tracking detector covers the pseudorapidity range \(|\eta| < 2.5\).
It consists of silicon pixel, silicon microstrip, and transition radiation tracking detectors.
Lead/liquid-argon (LAr) sampling calorimeters provide electromagnetic (EM) energy measurements
with high granularity.
A steel/scintillator-tile hadron calorimeter covers the central pseudorapidity range (\(|\eta| < 1.7\)).
The endcap and forward regions are instrumented with LAr calorimeters
for both the EM and hadronic energy measurements up to \(|\eta| = 4.9\).
The muon spectrometer surrounds the calorimeters and is based on
three large superconducting air-core toroidal magnets with eight coils each.
The field integral of the toroids ranges between \num{2.0} and \SI{6.0}{\tesla\metre}
across most of the detector.
The muon spectrometer includes a system of precision tracking chambers and fast detectors for triggering.
A two-level trigger system is used to select events.
The first-level trigger is implemented in hardware and uses a subset of the detector information
to accept events at a rate below \SI{100}{\kHz}.
This is followed by a software-based trigger that
reduces the accepted event rate to \SI{3}{\kHz} on average,
depending on the data-taking conditions.
An extensive software suite~\cite{ATL-SOFT-PUB-2021-001} is used in data simulation, in the reconstruction
and analysis of real and simulated data, in detector operations, and in the trigger and data acquisition
systems of the experiment.
 
\FloatBarrier

% End of text imported from the .//Detector.tex input file

% The next lines are included from the .//MCData.tex input file
\section{Data and MC samples}
\label{sec:datamc}
The analysis is performed using the data collected between August and October 2022 with the ATLAS detector, in $pp$ collisions at $\sqs = 13.6~\tev$.
After applying data-quality requirements~\cite{Aad:2019hmz}, the data sample corresponds to an integrated luminosity of $29~\ifb$.
The average number of inelastic $pp$ collisions in the dataset is about 42.
MC simulated samples are used in the analysis to optimise the event selection, estimate the selection efficiencies and correlation for the efficiencies of identifying and reconstructing two $b$-tagged jets, represented by the $C_b$ parameter, and predict contributions from various background processes.
If not stated otherwise, a top-quark mass of \(\mtop = \SI{172.5}{\GeV}\) is used, in the MC simulated samples and cross-section
calculations.
 
The production of \ttbar events was simulated using the \POWHEGBOX[v2]~\cite{Frixione:2007nw,Nason:2004rx,Frixione:2007vw,Alioli:2010xd}
generator at next-to-leading-order~(NLO) with the \NNPDF[3.0nlo]~\cite{Ball:2014uwa} PDF set and the \hdamp parameter set to 1.5\,\mtop~\cite{ATL-PHYS-PUB-2016-020}.\footnote{The \hdamp parameter is a resummation damping factor and one of the parameters that controls the matching of \POWHEG matrix elements to the parton shower and thus effectively regulates the high-\pT radiation against which the \ttbar system recoils.}
The events were interfaced to \PYTHIA[8.307]~\cite{Bierlich:2022pfr} to model the parton shower, hadronisation, and underlying event, with parameters set according to the A14 tune~\cite{ATL-PHYS-PUB-2014-021} and using the \NNPDF[2.3lo] set of PDFs~\cite{Ball:2012cx}.
The decays of bottom and charm hadrons were performed with \EVTGEN[2.1.1]~\cite{Lange:2001uf}.
The \ttbar sample was normalised to the cross-section prediction at next-to-next-to-leading-order~(NNLO) in QCD including the resummation of next-to-next-to-leading-logarithm~(NNLL) soft-gluon terms calculated using \TOPpp[2.0]~\cite{topxtheo1,topxtheo2,topxtheo3,topxtheo4,topxtheo5,topxtheo6,Czakon:2011xx}.
For proton--proton collisions at a centre-of-mass energy of \(\rts = \SI{13.6}{\TeV}\), this cross-section corresponds to \(\sigma(\ttbar)_{\text{NNLO+NNLL}}\) = $924^{+32}_{-40}$~pb, including all \ttbar decay channels, using the PDF4LHC21 PDF set~\cite{PDF4LHCWorkingGroup:2022cjn}.
The uncertainty includes variations in the renormalisation and factorisation scale, $\alpha_\text{s}$ and PDFs.
For the nominal prediction, the renormalisation and factorisation scales
were set equal to the top-quark mass, and the strong coupling constant was
set to $\alpha_\text{s}(m_Z)=0.118$. Samples of $\ttbar$ and $tW$ events with a different
top-quark mass have been simulated to estimate the variation of the
reconstruction efficiency as a function of this mass.
 
Events of $Z/\gamma^*\to\ellell$ production were simulated with the \SHERPA[2.2.12]~\cite{Bothmann:2019yzt} generator using NLO matrix elements~(ME) for up to two partons, and up to five partons at leading-order~(LO) calculated with the Comix~\cite{Gleisberg:2008fv} and \OPENLOOPS~\cite{Buccioni:2019sur,Cascioli:2011va,Denner:2016kdg} libraries.
They were matched with the \SHERPA parton shower~\cite{Schumann:2007mg} using the \MEPSatNLO prescription~\cite{Hoeche:2011fd,Hoeche:2012yf,Catani:2001cc,Hoeche:2009rj} using the set of tuned parameters developed by the \SHERPA authors.
The \NNPDF[3.0nnlo] set of PDFs was used.
Decays of $Z/\gamma^*\to \ee$, $Z/\gamma^*\to \mumu$ and $Z/\gamma^*\to\tautau$ were simulated.
The inclusive cross sections of the MC samples are normalised to NNLO in QCD and NLO in EW predictions, calculated using the MATRIX~\cite{Grazzini:2017mhc} program with the PDF4LHC21 PDF set.
In both the MC generator requirement and in the MATRIX calculation, leptons before QED radiation are used.
In the fiducial phase-space defined with lepton $\pt > 27~\gev$, $|\eta| < 2.5$ and $66 < m_{\ell\ell} < 116~\gev$, the predicted cross-section at NNLO in QCD and NLO in EW is $\sigma(Z)^{\mathrm{fid.}} = 746\pm 0.7\text{(integration)}^{+3}_{-5}(\text{scale})\pm21\text{(PDF)}$~pb for the $Z$-boson decaying into a single lepton flavour and leptons before QED radiation.\footnote{The integration uncertainty represents the uncertainty due to the finite number of points used in the MC integration of the phase-space integrals.}
 
For the calculation of the \ttbar\ to the $Z$-boson production cross-section ratio, several
PDF sets are considered: PDF4LHC21, CT18~\cite{Hou:2019efy}, CT18A~\cite{Hou:2019efy},
MSHT20~\cite{Bailey:2020ooq},
NNPDF4.0~\cite{Ball:2021leu}, ABMP16~\cite{Alekhin:2017kpj} and
ATLASpdf21~\cite{STDM-2020-32}. The uncertainties from the QCD scale variations are assumed to be uncorrelated between the processes, while the PDF variations are fully correlated in the calculation of the uncertainty.
The QCD scale uncertainty is evaluated by restricting the variations of the renormalisation and factorisation scales, where $\mu_\text{r}$ and $\mu_{\text{f}}$ are independently varied by a factor of two while meeting the requirement that they never differ from each other by more than a factor of two.
The predicted central values for the ratios for the different PDF sets vary between about 1.13 for the ABMP16 and 1.27 for the CT18 PDF set.
The uncertainties on the predictions range from about 4.5\% up to 9\%.
 
The associated production of single top quarks with \(W\) bosons (\(tW\)) was modelled with the \POWHEGBOX[v2] generator at NLO in QCD using the five-flavour scheme and the \NNPDF[3.0nlo] set of PDFs~\cite{Ball:2014uwa}.
The diagram-removal scheme~\cite{Frixione:2008yi} was used to remove interference and overlap with \ttbar production.
The events were interfaced to \PYTHIA[8.307] using the A14 tune and the \NNPDF[2.3lo] set of PDFs.
The inclusive cross-section was corrected to the theory prediction calculated at approximate N${}^{3}$LO in QCD using the PDF4LHC21 PDF set~\cite{Kidonakis:2021vob}.
The cross-section corresponds to \(\sigma(tW)_{\text{aN}^3\text{LO}} = 87.9^{+2.0}_{-1.9}\text{(scale)}\pm2.4\text{(PDF)}~\si{\pb}\).
 
Single-top-quark \(t\)-channel production was modelled with the \POWHEGBOX[v2] generator at NLO in QCD using the four-flavour scheme and the corresponding \NNPDF[3.0nlo] set of PDFs.
The events were interfaced with \PYTHIA[8.307] using the A14 tune and the
\NNPDF[2.3lo] set of PDFs.
Single-top-quark \(s\)-channel production was modelled using the \POWHEGBOX[v2]~\cite{Alioli:2009je,Nason:2004rx,Frixione:2007vw,Alioli:2010xd}
generator at NLO in QCD in the five-flavour scheme with the \NNPDF[3.0nlo]~\cite{Ball:2014uwa} PDF set.
The events were interfaced with \PYTHIA[8.307] using the A14
tune~\cite{ATL-PHYS-PUB-2014-021} and the \NNPDF[2.3lo] PDF set.
However, $t$-channel and $s$-channel productions contribute only via fake and non-prompt leptons.
 
The production of \(W+\)jets was simulated with the \SHERPA[2.2.12] generator with the same settings as the $Z$-boson sample.
The events were normalised to NNLO in QCD with NLO EW corrections calculated using the MATRIX software with the PDF4LHC21 PDF set.
 
Samples of diboson final states (\(VV\)) were simulated with the \SHERPA[2.2.12] generator, including off-shell effects and Higgs boson contributions where appropriate.
Fully-leptonic final states and semileptonic final states, where one boson decays leptonically and the other hadronically, were generated using
matrix elements at NLO accuracy in QCD for up to one additional parton and at LO accuracy for up to three additional parton emissions.
 
Samples of associated production of top-quark pairs with a $Z$-boson or a $W$-boson are simulated using the \SHERPA[2.2.12] generator.
Events with at least two leptons in the final state are considered.
 
The effect of multiple interactions in the same and neighbouring bunch crossings (pile-up) was modelled by overlaying~\cite{SIMU-2020-01} the original hard-scattering event with simulated inelastic $pp$ events generated by EPOS 2.0.1.4~\cite{Porteboeuf:2010um,Pierog:2013ria} and \PYTHIA~8.307.
The events generated with EPOS used the EPOS LHC tune, while the \PYTHIA~8 events used the NNPDF2.3 LO set of PDFs and parameter values set according to the A3 tune~\cite{ATL-PHYS-PUB-2016-017}.
\PYTHIA simulates events that contain at least one high transverse momentum (\pt) jet, isolated photon or lepton from the decay of $b$-hadrons, while EPOS simulates all remaining events in the sample.
The events are generated independently and a filter is applied to EPOS generated events to veto the events simulated by \PYTHIA.
The simulation has been reweighted to match the distribution of the number of primary vertices seen in data.
 
After the event generation, the ATLAS detector response is simulated by the toolkit \GEANT~\cite{Agostinelli:2002hh} with the full simulation of the ATLAS detector~\cite{SOFT-2010-01}.
The simulated samples are processed with the same software framework as the real data.
 
\FloatBarrier

% End of text imported from the .//MCData.tex input file

% The next lines are included from the .//Selection.tex input file
\section{Object definitions and event selection}
\label{sec:selection}
Events are selected if they include at least one lepton that is matched to a lepton identified by a single-lepton trigger.
The lowest \pt threshold in the single-lepton triggers is $26~\gev$ for electrons and $24~\gev$ for muons.
Events are also required to have at least one reconstructed collision vertex with two or more associated tracks
with \pt greater than 500~\mev.
The vertex with the highest $\sum \pt^2$ of the associated tracks is taken as the primary vertex.
 
Electron candidates are reconstructed from clusters of energy deposits in the electromagnetic calorimeter which are matched to particle tracks inside the ID.
The candidates need to pass the \emph{Tight} likelihood-based
identification criteria~\cite{EGAM-2018-01}.
Furthermore, they should pass the kinematic requirements $\pt > 27~\gev$ and $|\eta| < 2.47$, with the transition region between the barrel and the endcap calorimeters at $1.37 < |\eta| < 1.52$ excluded.
Additionally, electron candidates need to fulfil impact parameter
selection criteria: $|z_0 \sin{\theta}| < 0.5$~mm and $|d_0/\sigma(d_0)| <
5$.\footnote{$d_0$ is the distance of closest approach of the track to the primary vertex point, in the $r$---$\phi$ projection and $|z_0|$ is the $z$-coordinate of the track at the point of closest approach to the global $z$-axis.}
 
Muon candidates are reconstructed from tracks from the MS matched to tracks from the ID.
The candidates are required to pass \emph{Medium} identification
criteria~\cite{PERF-2015-10} as well as $\pt > 27~\gev$ and $|\eta| < 2.5$.
Additionally, muon candidates need to pass the selection of $|z_0 \sin{\theta}| < 0.5$~mm and $|d_0/\sigma(d_0)| < 3$.
 
To suppress non-prompt electrons or muons, isolation criteria are imposed.
The isolation is imposed by requiring the energy deposited in the
calorimeter in a cone $\Delta R < 0.2$ around the object, divided by the
object \pt, to be smaller than 0.15 and the momentum, reconstructed by the
ID, in a cone $\Delta R < 0.3$ around the object, divided by the object \pt, to be smaller than 0.04.
 
Dedicated scale factors (SFs) are used to correct the identification, reconstruction, isolation and trigger efficiencies to match the efficiencies in data.
For electrons, the identification SFs are calculated by reconstructing Run 2 data using the same software as used for Run 3 and following the methodology described in Refs.~\cite{EGAM-2018-01,PERF-2017-01}.
The nominal reconstruction and isolation SFs are assumed to be one, with the differences observed between data and MC efficiencies being covered by the uncertainties in the SFs.
For muons, the reconstruction, identification, isolation and track-to-vertex-association ($i.e.$ impact parameter requirements) efficiency SFs have been measured using the Run~3 data, using the methodology described in Ref.~\cite{MUON-2018-03}.
The electron (muon) momentum is corrected using calibrations obtained from Run~2 (Run~3) with conservative uncertainties.
Trigger SFs for both electron and muon triggers are measured in Run~3 in various data-taking periods and are applied to the simulation reflecting the relative luminosity of the individual periods.
 
Jets are reconstructed using a Particle Flow
algorithm~\cite{PERF-2015-09} that exploits both the calorimeter and ID
information. The \antikt~\cite{Cacciari:2008gp} jet algorithm is used with radius parameter $R=0.4$ implemented in the FastJet~\cite{Fastjet} software.
The calibration of jets uses simulation, representing the Run~3 conditions, to correct the jet-energy scale in both simulation and data.
Furthermore, the jet-energy scale in data is corrected using the \emph{in-situ} calibration measured in Run~2~\cite{JETM-2018-05}.
Any residual difference between Run~2 and Run~3 conditions after applying the calibration is considered as a systematic uncertainty estimated from the simulation.
After the calibration, jet candidates are required to have $\pt > 30~\gev$ and $|\eta| < 2.5$.
To suppress jets originating from pile-up, jets need to pass a neural-network-based Jet Vertex Tagger (NNJVT) discriminant for jets with \pt\ below 60~\gev. This is a successor of the Jet Vertex Tagger~\cite{PERF-2014-03}.
 
Jets containing $b$-hadrons are identified ($b$-tagged) using the DL1d algorithm.
The DL1d tagger uses a similar architecture as its predecessor DL1r, described in Ref.~\cite{FTAG-2019-07}, but employs
DeepSets~\cite{ATL-PHYS-PUB-2020-014} instead of a recurrent neural network for the processing of the track impact parameters.
The DL1d algorithm returns the probability of a jet being light-flavour, $c$-jet and $b$-jet, which is then combined into a single discriminant.
The jets are considered as $b$-tagged if the value of the discriminant passes the requirement corresponding to a 77\% efficiency for tagging a true $b$-jet.
The charm-jet (light-flavour-jet) rejection factor for jets with $\pt >
20\,\gev$ is about 6 (260) in \ttbar events.
Dedicated SFs are applied to the simulation to correct for the efficiency of tagging a true $b$-jet as a function of jet \pt.
No SFs are applied for the mis-tagging of $c$ and light-flavour jets, but their uncertainty is still considered.
To remove the ambiguity when one physical object is reconstructed as multiple objects, the overlap-removal algorithm used in Ref.~\cite{TOPQ-2016-01} is employed.
 
Selected events are required to have exactly two leptons (electrons or muons) of opposite electric charge.
In the same-flavour channels, events are required to have $66 <
m_{\ell\ell} < 116~\gev$.
No dilepton mass criteria are imposed on the $e\mu$ channel, but the events
are required to have one or two $b$-tagged jets, as described in Section~\ref{sec:strategy}.
 
Hadrons mis-identified as electrons as well as electrons from photon conversions and leptons from heavy flavour decays can be identified as prompt leptons.
These contributions are collectively referred to as fake or non-prompt leptons.
To estimate the contribution of fake or non-prompt leptons, for each simulated event, the MC particle-level information is checked to determine if the event that passed the selection contains a fake lepton or non-prompt lepton.
Based on this particle-level information, the events are split into real and fake or non-prompt lepton contributions.
The events classified as containing at least one fake or non-prompt lepton are combined into a single distribution, representing the estimate of this lepton contribution in data.
In the $e\mu$ channel, the expected contribution from the events with fake or non-prompt leptons accounts for about 1\% of all events, while in the same-flavour channels the contribution is below 0.1\% in both channels.
 
Figure~\ref{fig:DataMC1}(a) compares the multiplicity of $b$-tagged jets for data and prediction in the inclusive $e\mu$ channel.
The distribution of the leading lepton \pt in the $e\mu$ channel with at least
one $b$-tagged jet, which is dominated by \ttbar production, is shown in Figure~\ref{fig:DataMC1}(b).
Figures~\ref{fig:DataMC1}(c) and (d) compare the data and prediction for the leading lepton \pt distribution in the $ee$ channel and $\mu\mu$ channel, respectively, dominated by $Z$-boson production.
Figures~\ref{fig:DataMC1}(e) and (f) show the comparison of data and prediction for the invariant mass of the dilepton system in the $ee$ and $\mu\mu$ channel, respectively.
The observed \pt distribution of the selected leptons with the highest \pt\ in an event is harder than expected in the same-flavour channels and softer in the $e\mu$ channel.
However, in both cases the data agree with the prediction within the uncertainties indicated by the hashed band for the low-\pt regions that are relevant for this measurement as the lepton uncertainties impact the fitted cross section only via acceptance.
 
\begin{figure}[htbp]
\centering
\subfloat[]{
\includegraphics[width=0.49\textwidth]{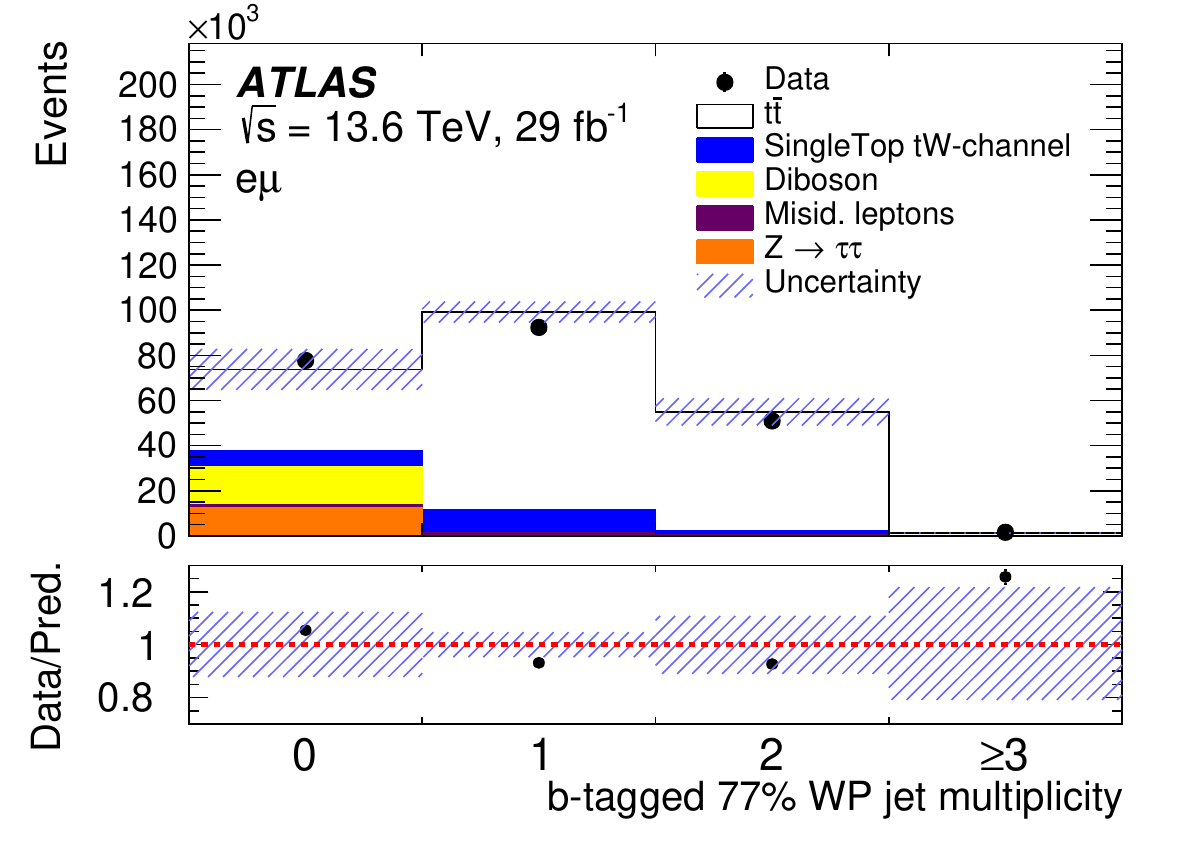}
}
\subfloat[]{
\includegraphics[width=0.49\textwidth]{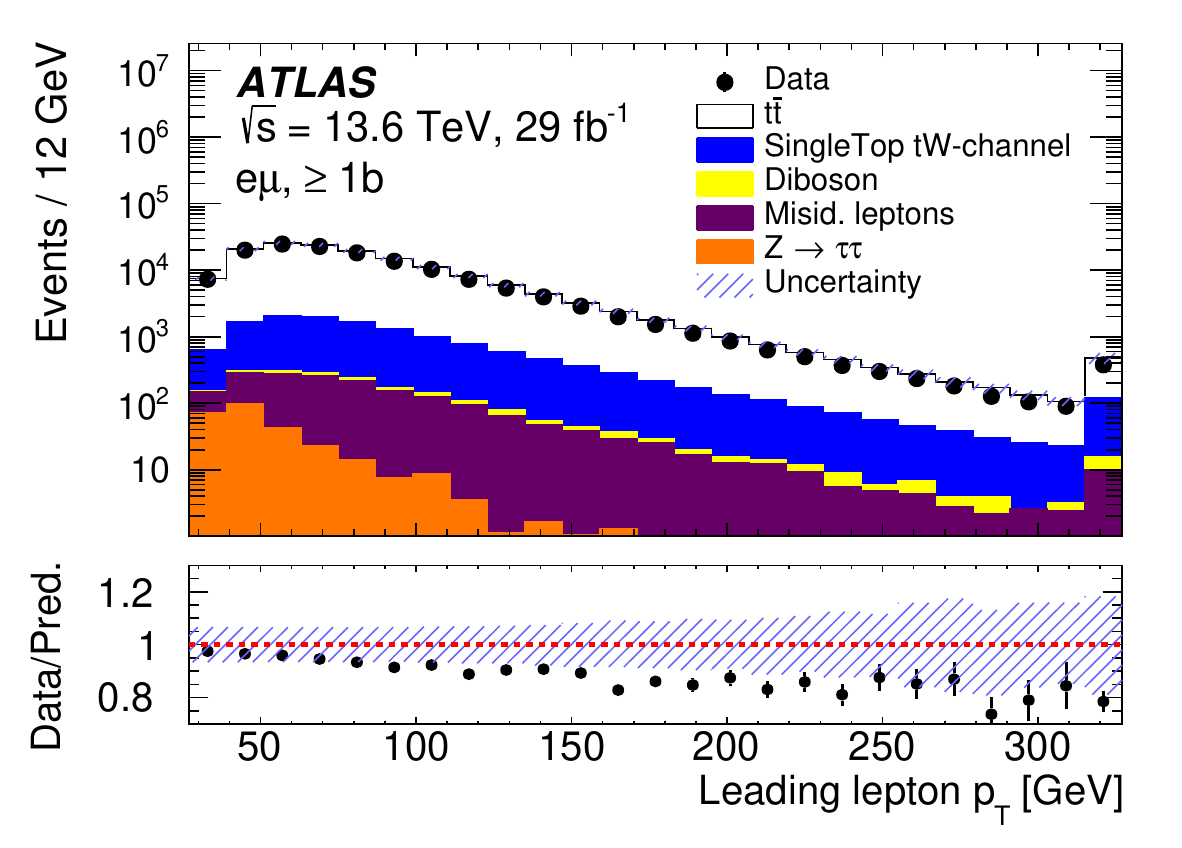}
} \\
\subfloat[]{
\includegraphics[width=0.49\textwidth]{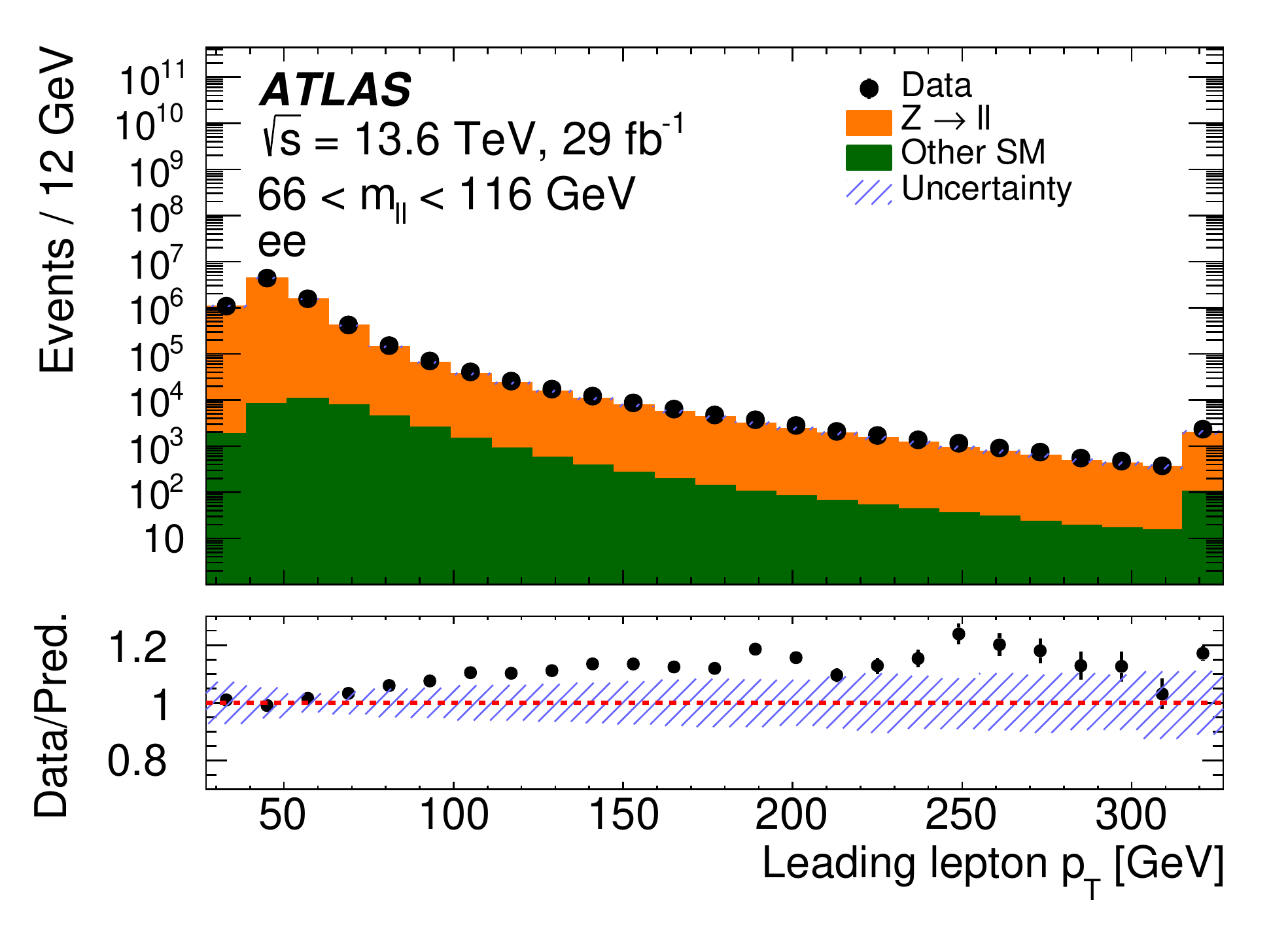}
}
\subfloat[]{
\includegraphics[width=0.49\textwidth]{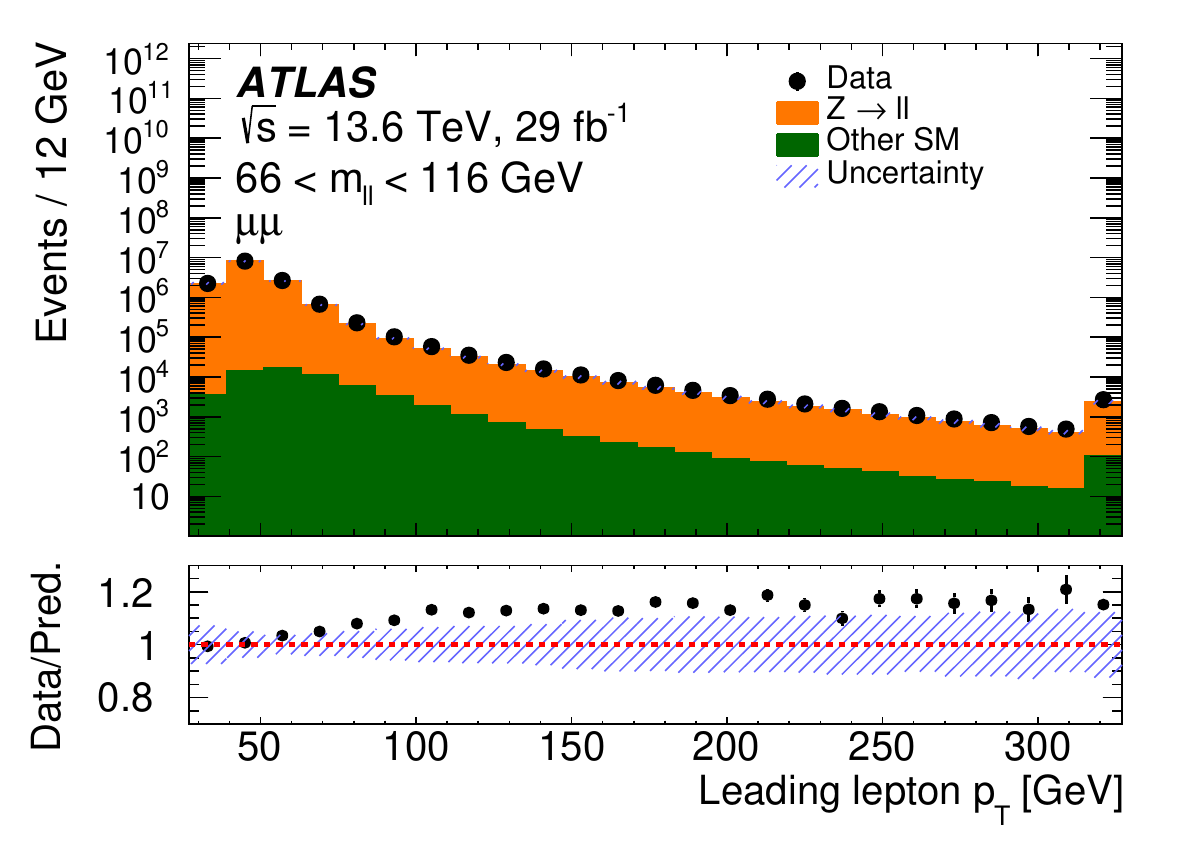}
} \\
\subfloat[]{
\includegraphics[width=0.49\textwidth]{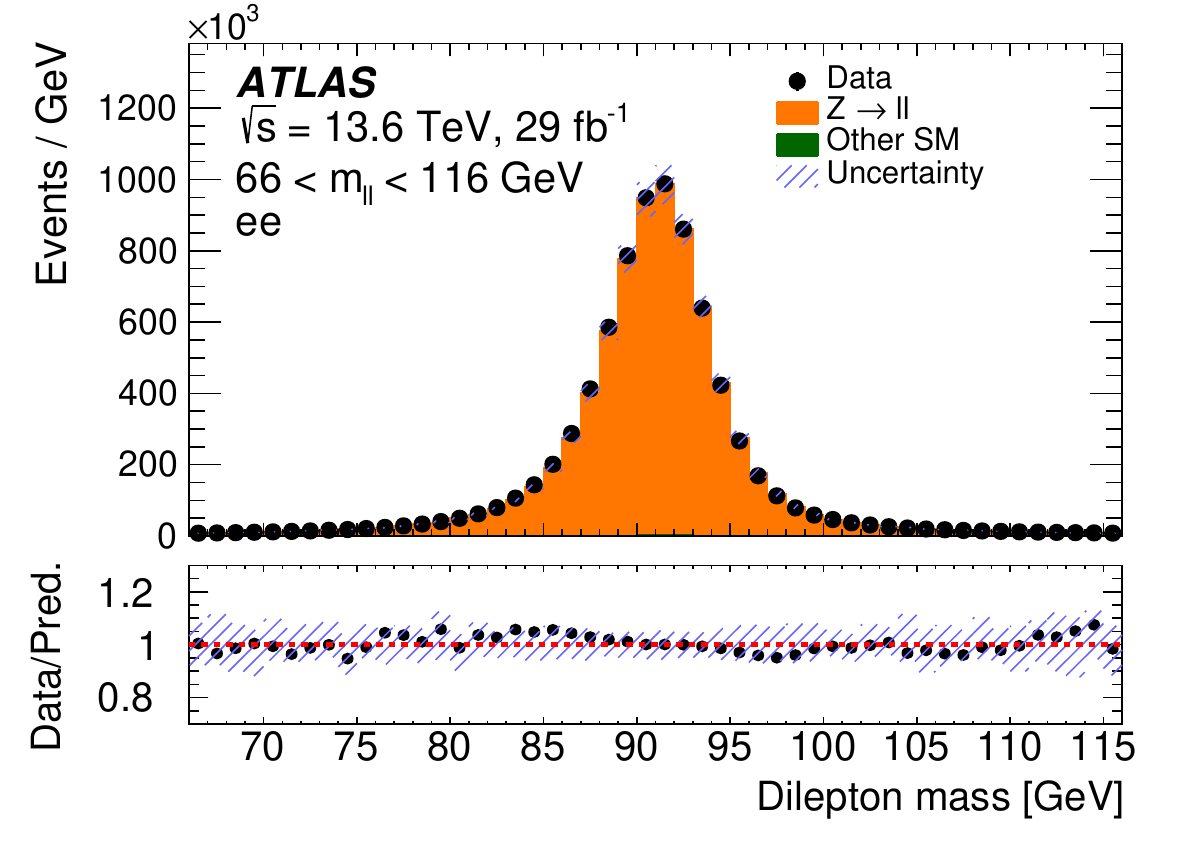}
}
\subfloat[]{
\includegraphics[width=0.49\textwidth]{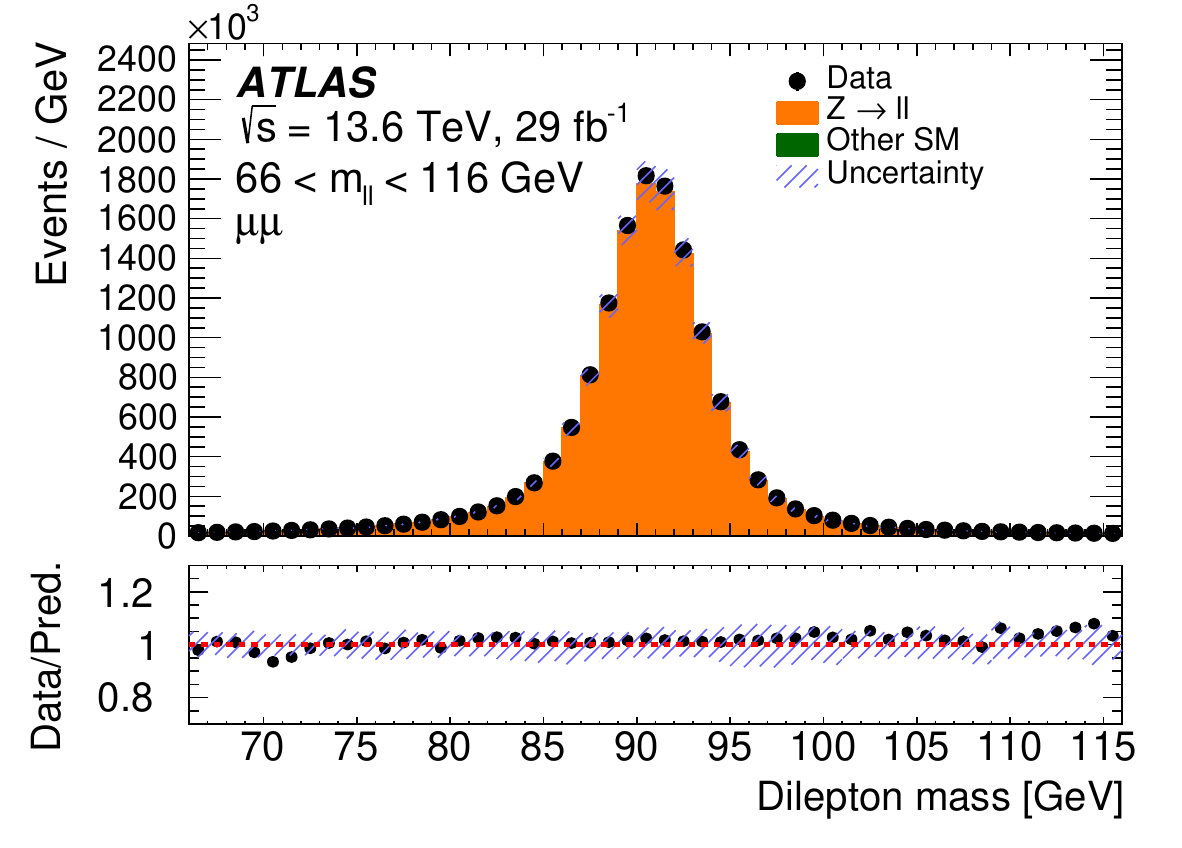}
}

\caption{Comparison of observed data and predictions for (a) the distribution
of the number of $b$-tagged jets in the inclusive $e\mu$ channel,
(b) the \pt\ of the leading lepton in the $e\mu$ channel with at least one
$b$-tagged jet,
(c) the leading lepton~\pt in the $ee$ channel, (d) the leading lepton~\pt in the $\mu\mu$ channel,
the invariant mass of the dilepton system in (e) the $ee$ channel and (f)
the $\mu\mu$ channel. The expected yields are calculated by normalising
the MC prediction to the product of integrated luminosity and predicted
cross section for each process. The ``Other SM'' represents all non-$Z$-boson SM processes merged into a single distribution.
The ``Misid. leptons'' label represents fake or
non-prompt leptons.  The hashed band represents the total
uncertainty. The bottom panel shows the ratio of data to the
prediction. The rightmost bins contain the overflow events.}
\label{fig:DataMC1}
\end{figure}
 
Table \ref{tab:DilepYields} shows the predicted and measured event yields
in the different dilepton categories.
The $e\mu$ inclusive region is only used in the simulation to calculate $C_b$.

\begin{table}[hbpt]
\caption{Predicted and observed event yields in the different dilepton regions
after event selection. Systematic uncertainties, representing all considered sources, are shown for the predicted yields.}
\centering
\resizebox{\textwidth}{!}{
\begin{tabular}{lr@{$\,\pm\,$}lr@{$\,\pm\,$}lr@{$\,\pm\,$}lr@{$\,\pm\,$}lr@{$\,\pm\,$}l}
\toprule
\textbf{Process} & \multicolumn{10}{c}{\textbf{Region}} \\
\midrule
\empty & \multicolumn{2}{c}{$e\mu$ inclusive} & \multicolumn{2}{c}{$e\mu$ 1 $b$} & \multicolumn{2}{c}{$e\mu$ 2 $b$} & \multicolumn{2}{c}{$ee$} & \multicolumn{2}{c}{$\mu\mu$} \\
\midrule
$\ttbar$ &       $177\,000$ & $5\,000$ & $87\,400$ & $2\,600$ & $52\,700$ & $1\,600$ & $22\,300$ & $1\,200$ & $33\,500$ & $1\,900$ \\
$Z\rightarrow ll$ &   $13\,100$ & $600$ & $270$ & $70$ & $14.2$ & $3.0$ & $7\,7490\,000$ & $230\,000$ & $14\,100\,000$ & $400\,000$ \\
SingleTop $tW$-channel & $19\,000$ & $9\,000$ & $10\,200$ & $700$ & $1\,720$ & $290$ & $2\,250$ & $120$ & $3\,410$ & $200$ \\
Diboson &            $17\,000$ & $8\,000$ & $230$ & $120$ & $5.3$ & $3.0$ & $9\,000$ & $5\,000$ & $18\,000$ & $9\,000$ \\
$t\bar{t}V$ &     $220$ & $110$ & $100$ & $50$ & $69$ & $35$ & $240$ & $120$ & $250$ & $130$ \\
Fake/non-prompt leptons & $2\,800$ & $1\,400$ & $1\,100$ & $500$ & $500$ & $500$ & $6\,000$ & $6\,000$ & $6\,000$ & $6\,000$ \\
\midrule
Total prediction & $229\,000$ & $14\,000$ & $99\,200$ & $3\,000$ & $55\,000$ & $1\,700$ & $7\,780\,000$ & $230\,000$ & $14\,100\,000$ & $400\,000$ \\
\midrule
Data & \multicolumn{2}{c}{$222\,711$} & \multicolumn{2}{c}{$92\,385$} & \multicolumn{2}{c}{$50\,956$} & \multicolumn{2}{c}{$7\,812\,978$} & \multicolumn{2}{c}{$14\,242\,875$} \\
\bottomrule
\end{tabular}
}
\label{tab:DilepYields}
\end{table}

% End of text imported from the .//Selection.tex input file

% The next lines are included from the .//Systematics.tex input file
\section{Systematic uncertainties}
\label{sec:systematics}
 
The uncertainty in the integrated luminosity for data recorded in 2022 is 2.2\%~\cite{PUB-LUMI-2023-01}, following the methodology discussed in Ref.~\cite{DAPR-2021-01}, using the LUCID-2 detector~\cite{LUCID2}
for the primary luminosity measurements, complemented by measurements using the inner detector and calorimeters.
The uncertainty is dominated by effects from horizontal-vertical correlations, known as non-factorisation, in the bunch-density distributions during the beam-separation scans used to determine the absolute luminosity scale, as well as by uncertainties in transferring this
calibration from the data-taking conditions of the absolute luminosity calibration to the conditions of routine physics data-taking.
Various beam-, calibration-method-, consistency-, reproducibility- and stability-related effects provide smaller contributions to the overall uncertainty.
The uncertainty in the pile-up modelling is estimated by varying the average number of interactions per bunch-crossing by 4\% in the simulation.
 
For the electron energy corrections, uncertainties obtained from the Run~2 calibrations using $Z\to ee$, as described in Ref.~\cite{EGAM-2018-01}, are increased,
covering the differences in the reconstruction between Run~2 and Run~3 of the LHC as estimated from the simulation.
The uncertainty in the muon momentum correction is estimated by separately varying the ID and MS components as estimated from $Z\to\mu\mu$ events, similarly to Ref.~\cite{PERF-2015-10}.
The uncertainties in the corrections are split into several components both for electrons and muons.
 
The uncertainties in the electron identification SFs are computed comparing simulation between Run~2 and Run~3.
It was checked that the SFs in Run 3 are within the estimated uncertainties in Run 2.
The uncertainties on the muon SFs have been calculated using the methods described in Ref.~\cite{MUON-2018-03}.
The trigger efficiencies with their uncertainties are measured using the
Run 3 data.
Most events can be triggered by either lepton, reducing the systematic
uncertainty from the modelling of the trigger efficiencies.
 
The lepton efficiencies (trigger, identification, reconstruction and
isolation) are generally the dominant detector uncertainties on the measurements. They depend on the lepton \pt, $\eta$ and $\phi$.
In the calculation of the uncertainty in \ratio, the nuisance parameters (NPs) corresponding to each systematic uncertainty for the detector related uncertainties are taken to be fully correlated between the \ttbar and $Z$-boson processes.
Correlating or treating the lepton isolation uncertainties separately between the \ttbar and $Z$-boson processes has no significant impact on the central value nor the total uncertainty in the \ratio\ fit.
 
The uncertainties in the jet-energy scale (JES)~\cite{PERF-2016-04} and jet-energy resolution (JER)~\cite{ATL-PHYS-PUB-2015-015} are considered, as estimated in Run~2.
Additionally, dedicated uncertainties due to the differences in reconstruction between Run~2 and Run~3 are considered; these are estimated from simulation.
A 10\% uncertainty per $b$-tagged jet is considered for events in the $e\mu$ channel to cover the discrepancies between data and simulation for the NNJVT discriminant.
The uncertainty is not applied to simulated \ttbar  events, as the jet reconstruction and tagging probability is  eventually determined in data as the $\epsilon_b$ parameter.
 
The $b$-tagging efficiencies for the $tW$ process and other backgrounds are estimated from the simulation.
The uncertainty in the $b$-jet efficiency, binned in jet \pt, is estimated using dilepton \ttbar\ events, employing a simplified version of the method described in Ref.~\cite{FTAG-2018-01}.
Separate, \pt-independent uncertainties are assigned to the charm- and light-flavours for the efficiency of \mbox{(mis-)tagging}.
These are 20\% for the $c$-mis-tag efficiency and 40\% for the light-flavour-mis-tag efficiency.
The uncertainties cover the differences in the modelling of the discriminant between data and simulation in \ttbar lepton+jets events for the charm mis-tag rate and in events with a $Z$-boson and at least one jet for the light-flavour mis-tag rate.
 
The impact of using a different parton shower and hadronisation model is evaluated
by comparing the nominal \ttbar sample with another sample produced by interfacing the
\POWHEGBOX[v2]~\cite{Frixione:2007nw,Nason:2004rx,Frixione:2007vw,Alioli:2010xd}
generator using the \NNPDF[3.0nlo]~\cite{Ball:2014uwa} PDF set with \HERWIG[7.2.3]~\cite{Bahr:2008pv,Bellm:2015jjp,Bellm:2019zci}.
Uncertainties due to the modelling of the \ttbar signal are estimated considering independent variations of the renormalisation and
factorisation scales in the ME of the nominal \ttbar sample by factors of 0.5 and 2, excluding variations that differ by a factor of four, but normalising the signal to the nominal cross section.
Additionally, variations of the renormalisation and factorisation scales are considered for the parton shower simulation affecting the initial- and final-state radiation, where the former is represented by the variation of the Var3c parameter of the A14 tune~\cite{ATL-PHYS-PUB-2017-007}.
The uncertainty due to the choice of the \hdamp\ parameter is estimated by increasing it by a factor of two.
To account for the uncertainty due to the top-quark \pt\ distribution mismodelling~\cite{TOPQ-2018-17}, the nominal \ttbar sample is compared with the same sample reweighted to match the top-quark \pt distribution computed at NNLO in QCD precision using the MATRIX program.
 
Several sources of systematic uncertainties in the modelling of the $Z$-boson events are considered.
Independent renormalisation and factorisation scale variations in the ME as well as the simultaneous variations of the ME and the parton shower by a factor of 0.5 and 2, excluding variations that differ by a factor of four, are tested.
The variation with the maximum impact on the yields is symmetrised and used to estimate the uncertainty in the analysis.
This variation originates from the simultaneous variation of the renormalisation and factorisation scale by a factor of 0.5 in the ME and the parton shower.
For all modelling uncertainties in the $Z$-boson production, the systematic variation is rescaled to have the same cross section in the fiducial phase-space.
 
A normalisation uncertainty of 3.5\% is assigned to the $tW$ process, reflecting the theoretical uncertainty in the cross section.
Similarly to \ttbar uncertainties, scale variations in the ME, Var3c and for the final-state radiation of the parton shower are considered for the single-top quark processes.
Furthermore, an uncertainty due to the interference between the $tW$ and \ttbar processes is obtained by comparing the diagram-removal scheme with the diagram-subtraction scheme~\cite{Frixione:2008yi}.
Scale variations between the \ttbar and $tW$ production processes are assumed to be uncorrelated, but this assumption is found to have no significant impact on the total expected uncertainty.
 
The PDF uncertainty for \ttbar and $Z$ modelling is estimated by considering the internal variations of the PDF4LHC21 PDF set~\cite{PDF4LHCWorkingGroup:2022cjn}.
This uncertainty is correlated between processes.
Only the acceptance effects are considered for the PDF and the modelling uncertainties.
 
A normalisation uncertainty of 50 (100)\% is assigned for the
fake and non-prompt lepton background in the $e\mu$ channel for events with
1 (2) $b$-tagged jets as estimated from the simulation, similarly to Ref.~\cite{TOPQ-2015-09}.
An uncertainty of 100\% is assigned on the normalisation of the fake and non-prompt lepton background in the $Z\to ee$ and $Z\to\mu\mu$ events.
A 50\% uncertainty is considered for the normalisation of the diboson processes and processes with associated production of the top-quark pairs with a $Z$-boson or a $W$-boson.
These normalisation uncertainties have no significant impact on the total uncertainty of the measurement.
 
\FloatBarrier

% End of text imported from the .//Systematics.tex input file

% The next lines are included from the .//Results.tex input file
\section{Results}
\label{sec:results}
 
Figure~\ref{fig:PrePost} shows the distributions used in the fit to extract
the \ttbar cross section and the ratio of the cross sections.
A profile-likelihood fit is used to extract \xtt\ and \ratio\ together with the
$b$-tagging efficiency $\epsilon_b$. The results of the fit are:
\begin{align*}
\xtt &=850 \pm 3\mathrm{(stat.)}\pm 18\mathrm{(syst.)}\pm
20\mathrm{(lumi.)~pb}, \\
\ratio &=1.145 \pm 0.003\mathrm{(stat.)}\pm 0.021\mathrm{(syst.)}\pm 0.002\mathrm{(lumi.)},\\
\epsilon_b &= 0.544 \pm 0.001 \mathrm{(stat.)} \pm 0.004 \mathrm{(syst.)} \pm 0.001 \mathrm{(lumi.)}.
\end{align*}
 
where the $Z$-boson cross section in $\ratio$ is measured in a fiducial phase-space defined by lepton $\pt > 27~\gev$, lepton $|\eta| < 2.5$ and an invariant mass of $66 < m_{\ell\ell} < 116~\gev$ for $\ell = e, \mu$.
 
This procedure takes into account all correlations among systematic
uncertainties in the \ttbar\ and $Z$-boson cross section measurements.
The obtained values for the parameters \xtt\ and \ratio\ have a positive correlation of about 0.6.
The result for \xtt\ is reported for a fixed top quark mass of $m_t = 172.5$~\gev,
and depends on the assumed value according to $(1/\xtt)\mathrm{d}\xtt/\mathrm{d}m_t=-0.36\%/\gev$,
as estimated from a variation of the top-quark mass in the \ttbar and $tW$ samples to 171.5~\gev and 173.5~\gev.
 
The predicted values for the $\ttbar$ cross section and the ratio \ratio\
extrapolated to the full phase assuming a top-quark mass of $172.5~\gev$ and
using the PDF4LHC21 PDF set are:
 
\begin{align*}
\sigma_{\ttbar}^{\text{theory}} &=924 ^{+32}_{-40}~(\text{scale+PDF+}\alpha_s)~\text{pb}, \\
R_{t\bar{t}/Z}^{\text{theory}} &= 1.238^{+0.063}_{-0.071}~(\text{scale+PDF+}\alpha_\text{s}).
\end{align*}
 
The $\epsilon_b$ value in the \ttbar\ simulation is $0.5431\pm 0.0003$~(MC stat.), compatible
within the per mille uncertainty with the fitted value.
No significant pulls or constrains of the NPs are present, with the exception of a NP for the uncertainty in the isolation of muons,
that is constrained to about 75\% of its pre-fit uncertainty.
The low constraining power in the fit is due to the presence of three
unconstrained parameters in the likelihood.
 
Figure \ref{fig:results1} shows the measured $\ttbar$ cross section as a function of the
centre-of-mass energy compared to the theory prediction using the PDF4LHC21 PDF set.
The $13.6$~\tev measurement is slightly lower than the prediction but compatible within 1.5 standard deviations, when adding the downward uncertainty on the prediction and upward measurement uncertainties in quadrature.
In Figure~\ref{fig:results2}, the ratio of the $\ttbar$ and $Z$-boson production cross sections is compared to
the prediction for several PDF sets.
The measured ratio is compatible with the SM prediction based on the PDF4LHC21 set at 1.3 standard deviations assuming a top-quark mass of 172.5~\gev.
 
To understand the cancellation of the systematic uncertainties, the
$Z$-boson cross section has been obtained from a fit to same-flavour events only. The result of the fit is:
 
\begin{align*}
\xz &= 744 \pm 11~\mathrm{(stat.+syst.)}\pm 16~\mathrm{(lumi.)~pb},
\end{align*}
 
using the same fiducial phase-space as was used for $\ratio$.
The statistical uncertainty is negligible compared to the systematic uncertainty.
 
In the $Z$-boson cross section fit, a supplementary normalisation uncertainty of 5.1\% is applied
on the \ttbar sample, according to the theoretical uncertainty on the
predicted cross section value.  The only pull and constraint on nuisance
parameters can come from the requirement that the $Z \rightarrow ee$ and
$Z \rightarrow \mu\mu$ cross sections are equal. Since these two
cross section measurements agree to better than 0.5 standard deviations, no
pulls are observed.
 
The theory prediction for the fiducial $Z \rightarrow \ell \ell$ cross section
is
 
\begin{align*}
\sigma_{Z\to\ell\ell}^{\text{fid.}, \text{theory}} &= 746^{+21}_{-22}~(\text{scale+PDF+}\alpha_\text{s})~\text{pb},
\end{align*}
agreeing with the measurement within 1 standard deviation.
Table~\ref{tab:zratio_GroupedImpact} shows the impact of the systematic uncertainties grouped by their origin.
The largest source of systematic uncertainty in the individual cross sections originates from the uncertainty in the luminosity estimation, followed by electron and muon reconstruction.
For the \ttbar\ cross section, the uncertainty in the parton shower and hadronisation modelling also has a significant impact on the precision.
For the cross section ratio, several sources of the systematic uncertainties partially cancel out.
The dominant uncertainty for the cross section ratio originates from \ttbar\ modelling, as it does not cancel out.
Other significant uncertainties on the ratio include uncertainty in the trigger efficiencies and lepton reconstruction efficiencies.
The trigger efficiency uncertainty does not fully cancel out as there are different $\eta$ and \pt\ distributions in \ttbar and $Z$-boson events, and different combinations of triggers employed.
 
The simulated distribution of the \pt\ of the dilepton pair is known to not be accurately modelled by the MC simulation~\cite{PMGR-2021-01}.
As a cross check, the simulated distribution of the \pt\ of the dilepton pair has been reweighted to match the data in $ee$ and $\mu\mu$ events, assuming that only $Z$-boson events contribute.
After the reweighting, the impact on the total predicted yields was found to be below 0.13\%, thus negligible.
 
\begin{figure}[htb]
\centering
\includegraphics[width=0.49\textwidth]{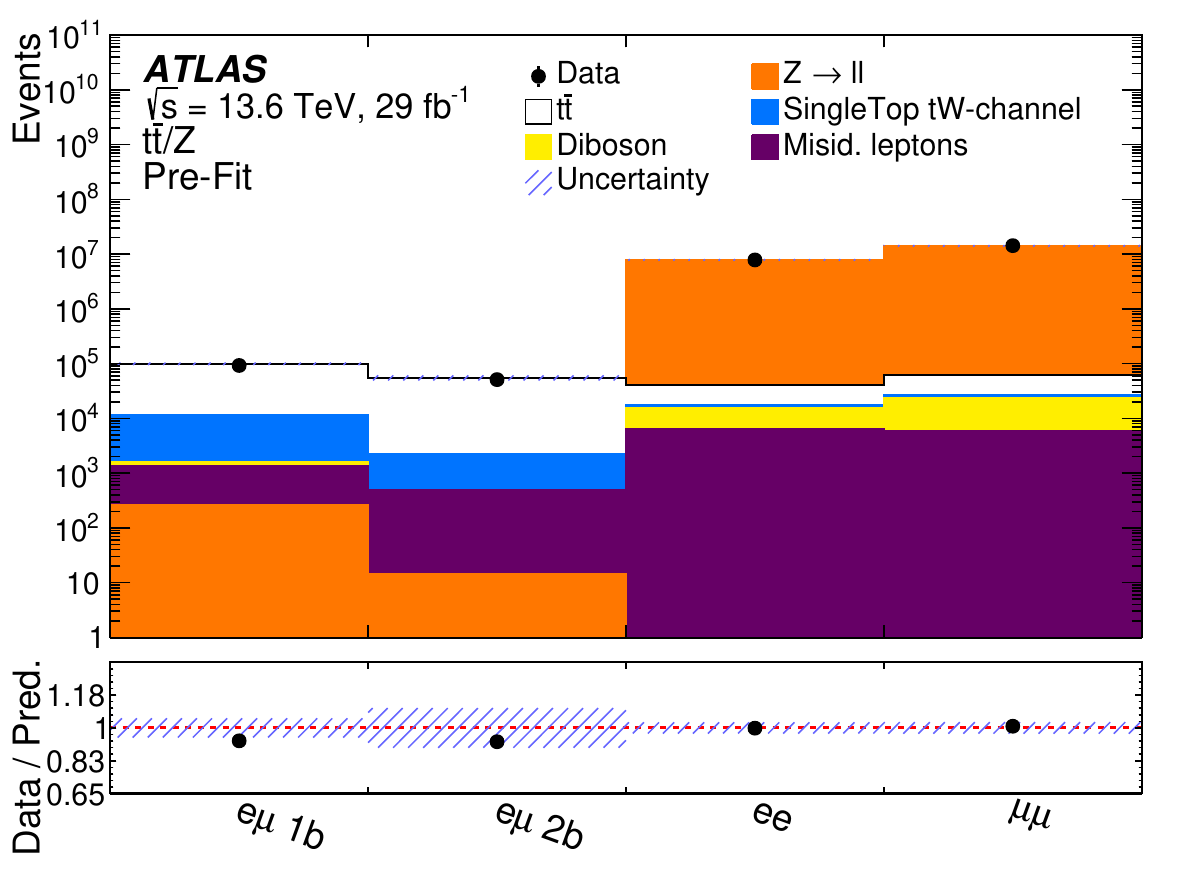}
\includegraphics[width=0.49\textwidth]{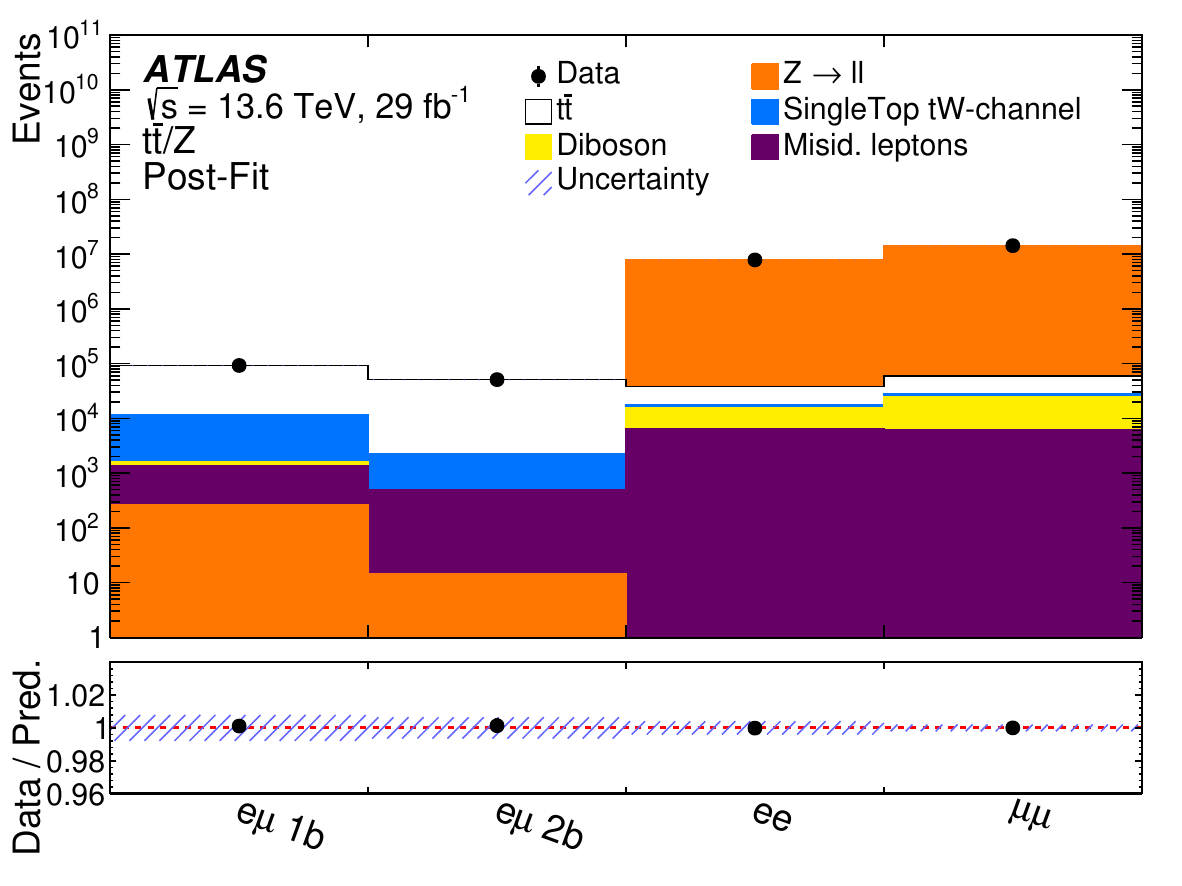}
\caption{Comparison of data and prediction for the event yields before
(left) and after the fit (right) in the regions used in the $\ttbar$ fit.
The bottom panel shows the ratio of data over the prediction. The
hashed bands represent the total uncertainty. Correlations of the NPs
as obtained from the fit are used to build the uncertainty band in the
post-fit distribution. The error bars on the data points are smaller than the displayed points.}
\label{fig:PrePost}
\end{figure}
 
\begin{table}[!hbt]
\centering
\caption{ Observed impact of the different sources of uncertainty on
the $\ttbar$ and $Z$-boson cross sections and their ratio $R$,
grouped by category. The $\ttbar$ cross section and $R$ are obtained
in a simultaneous fit to the four regions shown in Figure~\ref{fig:PrePost},
while the $Z$-boson cross section is obtained from
a fit to same-flavour events only. The impact of each category is
obtained by repeating the fit after having fixed the set of nuisance
parameters corresponding to that category, and subtracting that
uncertainty in quadrature from the uncertainty found in the full
fit. The statistical uncertainty is obtained by repeating the fit after having fixed all
nuisance parameters to their fitted values. Only the acceptance
effects are considered for the PDF and the modelling uncertainties.}
\begin{tabular}{llccc}
\toprule
& \textbf{Category} & \multicolumn{3}{c}{\textbf{Uncertainty} [\%]}       \\
\toprule
&  &  $\sigma_{t\bar{t}}$  & $\sigma_{Z \to\ell\ell}^{\text{fid.}}$ & $R_{t\bar{t}/Z}$ \\
\midrule
$t\bar{t}$    & $t\bar{t}$ parton shower/hadronisation    & 0.9  & $<0.2$ & 0.9 \\
& $t\bar{t}$ scale variations                 & 0.4  & $<0.2$ & 0.4 \\
& $t\bar{t}$ normalisation                    & -  & $<0.2$ & - \\
& Top quark \pt reweighting                   & 0.6  & $<0.2$ & 0.6 \\
$Z$       & $Z$ scale variations                        & $<0.2$  & 0.4 & 0.3 \\
Bkg.        & Single top modelling                        & 0.6  & $<0.2$ & 0.6 \\
& Diboson modelling                           & $<0.2$  & $<0.2$ & $0.2$ \\
& $t\bar{t}V$ modelling                       & $<0.2$  & $<0.2$ & $<0.2$ \\
& Fake and non-prompt leptons                 & 0.6  & $<0.2$ & 0.6 \\
Lept.       & Electron reconstruction                     & 1.2  & 1.0 & 0.4 \\
& Muon reconstruction                         & 1.4  & 1.4 & 0.3 \\
& Lepton trigger                              & 0.4  & 0.4 & 0.4 \\
Jets/tagging & Jet reconstruction                         & 0.4  & - & 0.4 \\
& Flavour tagging                             & 0.4  & - & 0.3 \\
\midrule
& PDFs                                        & 0.5  & $<0.2$ & 0.5 \\
& Pileup                                      & 0.7  & 0.8 & $<0.2$ \\
& Luminosity                                  & 2.3  & 2.2 & 0.3 \\
\bottomrule
& Systematic uncertainty                      & 3.2  & 2.8 & 1.8 \\
& Statistical uncertainty                     & 0.3  & 0.02 & 0.3 \\
\midrule
& Total uncertainty                           & 3.2  & 2.8 & 1.9 \\
\bottomrule
\end{tabular}
\label{tab:zratio_GroupedImpact}
\end{table}
 
\begin{figure}[htb]
\centering
\includegraphics[width=0.69\textwidth]{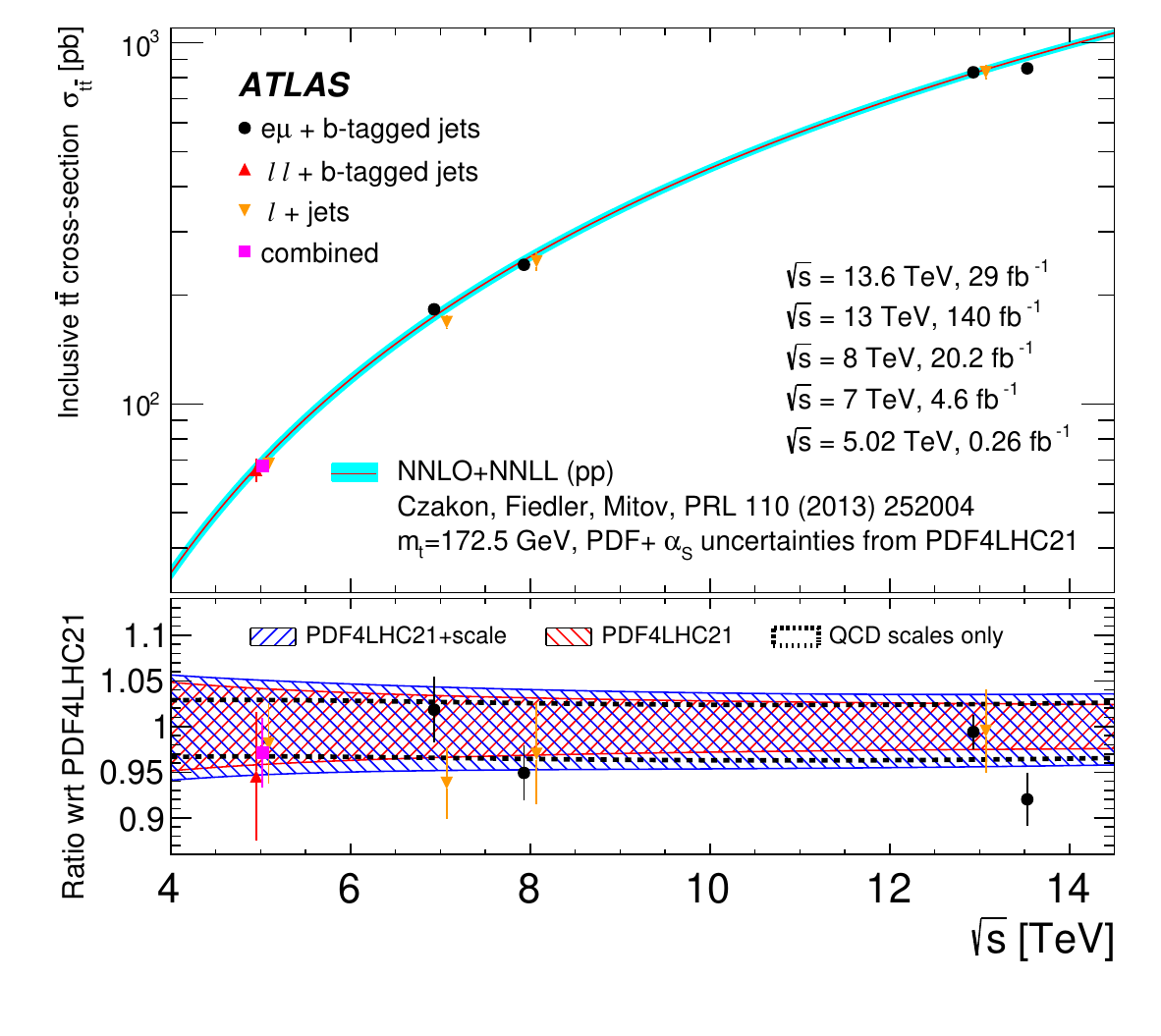}
\caption{Comparison of the measured $\ttbar$ cross sections at various centre-of-mass energies and the theory predictions using the PDF4LHC21 PDF set.
Measurements using the $e\mu$ final state ($e\mu$ + $b$-tagged jets); $ee$, $\mu\mu$ and $e\mu$ final states ($ll$ + $b$-tagged jets); single lepton final states ($l$ + jets); as well as combinations of final states (combined) are compared.
The bottom panel shows the ratio of the measured values and three predictions that either contain only the uncertainties originating from the QCD scale variations (black),
only the variations in the PDF uncertainties (red) or the total uncertainty in the prediction (blue).}
\label{fig:results1}
\end{figure}
 
\begin{figure}[htb]
\centering
\includegraphics[width=0.89\textwidth,trim={2cm 3cm 0 0},clip]{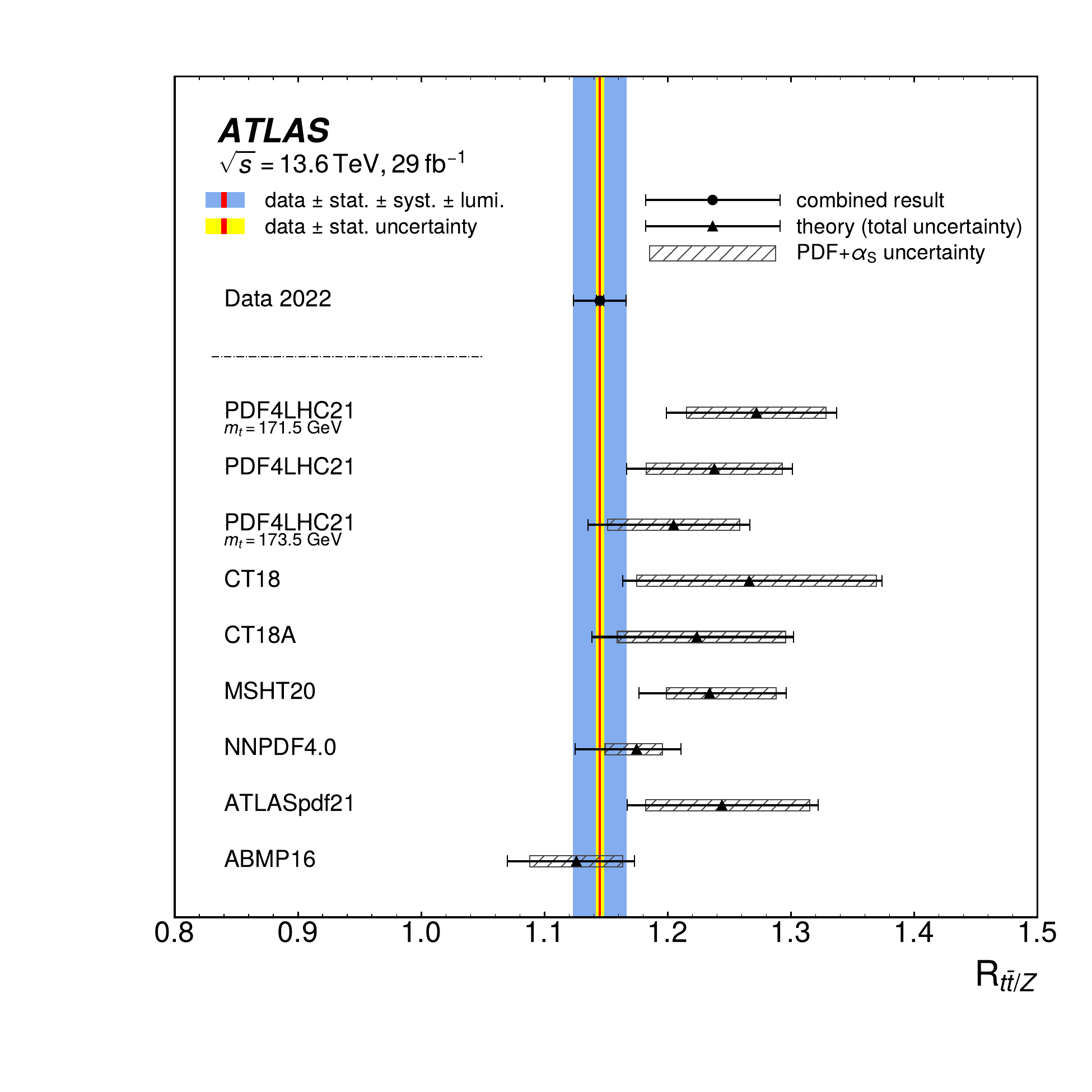}
\caption{Ratio of the $\ttbar$ to the $Z$-boson cross section
compared to the predictions for several sets of parton distribution
functions. The total cross section for \ttbar production and the fiducial cross section for $Z$-boson production are used.
If not explicitly mentioned, a top-quark mass of 172.5 \gev\ is assumed.
The hatched box represents the PDF+$\alpha_\text{S}$ component of the uncertainty on the prediction.
For the PDF4LHC21 PDF set, predictions for different assumptions about the top-quark mass are also displayed.}
\label{fig:results2}
\end{figure}
 
\FloatBarrier

% End of text imported from the .//Results.tex input file

% The next lines are included from the .//Conclusions.tex input file
\section{Conclusions}
\label{sec:conclusions}
The first ATLAS measurements of the inclusive \ttbar production cross
section and the ratio of the $\ttbar$ to the $Z$-boson cross section
at the new LHC centre-of-mass energy of 13.6\,\tev are presented,
using 29~$\ifb$ of data collected in 2022.
The \ttbar cross section is measured to be $\xtt =850 \pm 3\mathrm{(stat.)}\pm 18\mathrm{(syst.)}\pm 20\mathrm{(lumi.)}$~pb.
The measured value for the ratio is $\ratio = 1.145 \pm
0.003\mathrm{(stat.)}\pm 0.021\mathrm{(syst.)}\pm
0.002\mathrm{(lumi.)}$. Both measurements agree with the
Standard Model predictions of $924 ^{+32}_{-40}
(\text{scale+PDF+}\alpha_\text{s})~\text{pb}$ for the \ttbar cross section
and of $R_{t\bar{t}/Z}^{\text{theory}} = 1.238^{+0.063}_{-0.071}
(\text{scale+PDF+}\alpha_\text{s})$  for the ratio, using the PDF4LHC21 PDF set.
The absolute cross section measurement is limited by the uncertainty in the luminosity measurement as well as the lepton efficiency uncertainties.
In the ratio of the \ttbar\ and $Z$ cross sections, however, the luminosity uncertainty and lepton uncertainties cancel out to a large extent resulting in the total uncertainty of about 1.9\%.
A top-quark mass of 172.5~\gev is assumed for the prediction and for the simulation used to correct data.
The dependence on the Monte Carlo top-quark mass parameter on the cross section is provided.
 
\FloatBarrier

% End of text imported from the .//Conclusions.tex input file

\section*{Acknowledgements}

% The next lines are included from the .//acknowledgements/Acknowledgements.tex input file

We thank CERN for the very successful operation of the LHC, as well as the
support staff from our institutions without whom ATLAS could not be
operated efficiently.
 
We acknowledge the support of
ANPCyT, Argentina;
YerPhI, Armenia;
ARC, Australia;
BMWFW and FWF, Austria;
ANAS, Azerbaijan;
CNPq and FAPESP, Brazil;
NSERC, NRC and CFI, Canada;
CERN;
ANID, Chile;
CAS, MOST and NSFC, China;
Minciencias, Colombia;
MEYS CR, Czech Republic;
DNRF and DNSRC, Denmark;
IN2P3-CNRS and CEA-DRF/IRFU, France;
SRNSFG, Georgia;
BMBF, HGF and MPG, Germany;
GSRI, Greece;
RGC and Hong Kong SAR, China;
ISF and Benoziyo Center, Israel;
INFN, Italy;
MEXT and JSPS, Japan;
CNRST, Morocco;
NWO, Netherlands;
RCN, Norway;
MEiN, Poland;
FCT, Portugal;
MNE/IFA, Romania;
MESTD, Serbia;
MSSR, Slovakia;
ARRS and MIZ\v{S}, Slovenia;
DSI/NRF, South Africa;
MICINN, Spain;
SRC and Wallenberg Foundation, Sweden;
SERI, SNSF and Cantons of Bern and Geneva, Switzerland;
MOST, Taiwan;
TENMAK, T\"urkiye;
STFC, United Kingdom;
DOE and NSF, United States of America.
In addition, individual groups and members have received support from
BCKDF, CANARIE, Compute Canada and CRC, Canada;
PRIMUS 21/SCI/017 and UNCE SCI/013, Czech Republic;
COST, ERC, ERDF, Horizon 2020, ICSC-NextGenerationEU and Marie Sk{\l}odowska-Curie Actions, European Union;
Investissements d'Avenir Labex, Investissements d'Avenir Idex and ANR, France;
DFG and AvH Foundation, Germany;
Herakleitos, Thales and Aristeia programmes co-financed by EU-ESF and the Greek NSRF, Greece;
BSF-NSF and MINERVA, Israel;
Norwegian Financial Mechanism 2014-2021, Norway;
NCN and NAWA, Poland;
La Caixa Banking Foundation, CERCA Programme Generalitat de Catalunya and PROMETEO and GenT Programmes Generalitat Valenciana, Spain;
G\"{o}ran Gustafssons Stiftelse, Sweden;
The Royal Society and Leverhulme Trust, United Kingdom.
 
The crucial computing support from all WLCG partners is acknowledged gratefully, in particular from CERN, the ATLAS Tier-1 facilities at TRIUMF (Canada), NDGF (Denmark, Norway, Sweden), CC-IN2P3 (France), KIT/GridKA (Germany), INFN-CNAF (Italy), NL-T1 (Netherlands), PIC (Spain), ASGC (Taiwan), RAL (UK) and BNL (USA), the Tier-2 facilities worldwide and large non-WLCG resource providers. Major contributors of computing resources are listed in Ref.~\cite{ATL-SOFT-PUB-2023-001}.

% End of text imported from the .//acknowledgements/Acknowledgements.tex input file

\printbibliography

@Article{topxtheo1,
  author        = "Cacciari, Matteo and Czakon, Michal and Mangano, Michelangelo and Mitov, Alexander and Nason, Paolo",
  title         = "Top-pair production at hadron colliders with next-to-next-to-leading logarithmic soft-gluon resummation", 
  journal       = "Phys.\ Lett.\ B",
  doi     = {10.1016/j.physletb.2012.03.013},
  year          = "2012",
  volume        = "710",
  pages         = "612",
  archivePrefix = "arXiv",
  eprint        = "1111.5869",
  primaryClass  = "hep-ph"
}

@Article{topxtheo2,
  author        = "{P. B\"arnreuther, M. Czakon and A. Mitov}",
  title         = "Percent-Level-Precision Physics at the Tevatron: Next-to-Next-to-Leading Order QCD Corrections to $q\bar{q}\rightarrow t\bar{t}+X$",
 doi    = {10.1103/PhysRevLett.109.132001},
  journal       = "Phys.\ Rev.\ Lett.",
  year          = "2012",
  volume        = "109",
  pages         = "132001",
  archivePrefix = "arXiv",
  eprint        = "1204.5201",
  primaryClass  = "hep-ph"
}

@Article{topxtheo3,
  author        = "M. Czakon and A. Mitov",
  title         = "NNLO corrections to top-pair production at hadron colliders: the all-fermionic scattering channels", 
  journal       = "JHEP",
  doi = {10.1007/JHEP12(2012)054},
  year          = "2012",
  volume        = "12",
  pages         = "054",
  archivePrefix = "arXiv",
  eprint        = "1207.0236",
  primaryClass  = "hep-ph"
}

@Article{topxtheo4,
  author        = "M. Czakon and A. Mitov",
  title         = "NNLO corrections to top pair production at hadron colliders: the quark-gluon reaction", 
  journal       = "JHEP",
  doi   = {10.1007/JHEP01(2013)080},
  year          = "2013",
  volume        = "01",
  pages         = "080",
  archivePrefix = "arXiv",
  eprint        = "1210.6832",
  primaryClass  = "hep-ph"
}

@Article{topxtheo5,
  author        = "M. Czakon and P. Fiedler and A. Mitov",
  title         = "Total Top-Quark Pair Production Cross Section at Hadron Colliders Through $\mathcal{O}(\alpha_S^4)$", 
  journal       = "Phys.\ Rev.\ Lett.",
  doi     = {10.1103/PhysRevLett.110.252004},
  year          = "2013",
  volume        = "110",
  pages         = "252004",
  archivePrefix = "arXiv",
  eprint        = "1303.6254",
  primaryClass  = "hep-ph",
}

@article{topxtheo6,
      author         = "Catani, Stefano and Devoto, Simone and Grazzini,
                        Massimiliano and Kallweit, Stefan and Mazzitelli, Javier
                        and Sargsyan, Hayk",
      title          = "{Top-quark pair hadroproduction at
                        next-to-next-to-leading order in QCD}",
      journal        = "Phys. Rev. D",
      volume         = "99",
      year           = "2019",
      number         = "5",
      pages          = "051501",
      doi            = "10.1103/PhysRevD.99.051501",
      eprint         = "1901.04005",
      archivePrefix  = "arXiv",
      primaryClass   = "hep-ph",
      reportNumber   = "ZU-TH 02/19",
      SLACcitation   = "%%CITATION = ARXIV:1901.04005;%%"
}

@article{Mangano:2012mh,
    author = "Mangano, Michelangelo L. and Rojo, Juan",
    title = "{Cross Section Ratios between different CM energies at the LHC: opportunities for precision measurements and BSM sensitivity}",
    eprint = "1206.3557",
    archivePrefix = "arXiv",
    primaryClass = "hep-ph",
    reportNumber = "CERN-PH-TH-2012-159",
    doi = "10.1007/JHEP08(2012)010",
    journal = "JHEP",
    volume = "08",
    pages = "010",
    year = "2012"
}

@article{Rojo:2015acz,
    author = "Rojo, Juan and others",
    title = "{The PDF4LHC report on PDFs and LHC data: Results from Run I and preparation for Run II}",
    eprint = "1507.00556",
    archivePrefix = "arXiv",
    primaryClass = "hep-ph",
    reportNumber = "OUTP-15-11P, LCTS-2015-14, GLAS-PPE-2015-01, DESY-15-088, CERN-PH-TH-2015-150, JLAB-THY-15-2064",
    doi = "10.1088/0954-3899/42/10/103103",
    journal = "J. Phys. G",
    volume = "42",
    pages = "103103",
    year = "2015"
}

@book{Porteboeuf:2010um,
    author = "Porteboeuf, S. and Pierog, T. and Werner, K.",
    title = "{Producing Hard Processes Regarding the Complete Event: The EPOS Event Generator}",
    booktitle = "{45th Rencontres de Moriond on QCD and High Energy Interactions}",
    eprint = "1006.2967",
    archivePrefix = "arXiv",
    primaryClass = "hep-ph",
    publisher = "Gioi Publishers",
    pages = "135--140",
    year = "2010"
}

@article{Pierog:2013ria,
    author = "Pierog, T. and Karpenko, Iu. and Katzy, J. M. and Yatsenko, E. and Werner, K.",
    title = "{EPOS LHC: Test of collective hadronization with data measured at the CERN Large Hadron Collider}",
    eprint = "1306.0121",
    archivePrefix = "arXiv",
    primaryClass = "hep-ph",
    reportNumber = "DESY-13-125",
    doi = "10.1103/PhysRevC.92.034906",
    journal = "Phys. Rev. C",
    volume = "92",
    number = "3",
    pages = "034906",
    year = "2015"
}

@article{Grazzini:2017mhc,
    author = "Grazzini, Massimiliano and Kallweit, Stefan and Wiesemann, Marius",
    title = "{Fully differential NNLO computations with MATRIX}",
    eprint = "1711.06631",
    archivePrefix = "arXiv",
    primaryClass = "hep-ph",
    reportNumber = "ZU-TH-30-17, CERN-TH-2017-232, ZU-TH 30/17",
    doi = "10.1140/epjc/s10052-018-5771-7",
    journal = "Eur. Phys. J. C",
    volume = "78",
    number = "7",
    pages = "537",
    year = "2018"
}

@article{PDF4LHCWorkingGroup:2022cjn,
    author = "Ball, Richard D. and others",
    collaboration = "PDF4LHC Working Group",
    title = "{The PDF4LHC21 combination of global PDF fits for the LHC Run III}",
    eprint = "2203.05506",
    archivePrefix = "arXiv",
    primaryClass = "hep-ph",
    reportNumber = "Edinburgh 2021/31, FERMILAB-PUB-22-121-QIS-SCD-T, MSUHEP-22-010,
  Nikhef 2021-033, SMU-HEP-22-01",
    doi = "10.1088/1361-6471/ac7216",
    journal = "J. Phys. G",
    volume = "49",
    number = "8",
    pages = "080501",
    year = "2022"
}

@article{Aad:2019hmz,
    archiveprefix = {arXiv},
    author = {{ATLAS Collaboration}},
    doi = {10.1088/1748-0221/15/04/P04003},
    eprint = {1911.04632},
    journal = {JINST},
    number = {04},
    pages = {P04003},
    primaryclass = {physics.ins-det},
    reportnumber = {CERN-EP-2019-207},
    title = {{ATLAS data quality operations and performance for 2015\textendash{}2018 data-taking}},
    volume = {15},
    year = {2020}}

@article{LUCID2,
    author = "Avoni, G. and others",
    title = "{The new LUCID-2 detector for luminosity measurement and monitoring in ATLAS}",
    doi = "10.1088/1748-0221/13/07/P07017",
    journal = "JINST",
    volume = "13",
    number = "07",
    pages = "P07017",
    year = "2018"
}

@article{TOPQ-2018-40-fix,
    author         = "{ATLAS Collaboration}",
    title          = "{Measurement of the \(t\bar{t}\) production cross-section in \(pp\) collisions at \(\sqrt{s} = 5.02\,\text{TeV}\) with the ATLAS detector}",
    eprint = "2207.01354",
    archivePrefix = "arXiv",
    primaryClass = "hep-ex",
    reportNumber = "CERN-EP-2022-119",
    doi = "10.1007/JHEP06(2023)138",
    journal = "JHEP",
    volume = "06",
    pages = "138",
    year = "2023"
}

@misc{Gavin:2010az-fix,
     author    = "Gavin, Ryan and Li, Ye and Petriello, Frank and Quackenbush, Seth",
     title     = "{FEWZ 2.0: A code for hadronic \(Z\) production at next-to-next-to-leading order}",
     year      = "2010",
     pages   = "2388",
     eprint    = "1011.3540",
     archivePrefix = "arXiv",
     primaryClass  =  "hep-ph",
     SLACcitation  = "%%CITATION = 1011.3540;%%"
}

@article{Bierlich:2022pfr,
    author = "Bierlich, Christian and others",
    title = "{A comprehensive guide to the physics and usage of PYTHIA 8.3}",
    journal={SciPost Phys. Codebases},
    eprint = "2203.11601",
    archivePrefix = "arXiv",
    primaryClass = "hep-ph",
    reportNumber = "LU-TP 22-16, MCNET-22-04, FERMILAB-PUB-22-227-SCD",
    doi = "10.21468/SciPostPhysCodeb.8",
    month = "3",
    year = "2022"
}

@article{Kidonakis:2021vob,
    author = "Kidonakis, Nikolaos and Yamanaka, Nodoka",
    title = "{Higher-order corrections for $tW$ production at high-energy hadron colliders}",
    eprint = "2102.11300",
    archivePrefix = "arXiv",
    primaryClass = "hep-ph",
    doi = "10.1007/JHEP05(2021)278",
    journal = "JHEP",
    volume = "05",
    pages = "278",
    year = "2021"
}

@article{Hou:2019efy,
    author = "Hou, Tie-Jiun and others",
    title = "{New CTEQ global analysis of quantum chromodynamics with high-precision data from the LHC}",
    eprint = "1912.10053",
    archivePrefix = "arXiv",
    primaryClass = "hep-ph",
    reportNumber = "MSUHEP-19-025, PITT-PACC-1911, SMU-HEP-19-03",
    doi = "10.1103/PhysRevD.103.014013",
    journal = "Phys. Rev. D",
    volume = "103",
    number = "1",
    pages = "014013",
    year = "2021"
}

@article{Bailey:2020ooq,
    author = "Bailey, S. and Cridge, T. and Harland-Lang, L. A. and Martin, A. D. and Thorne, R. S.",
    title = "{Parton distributions from LHC, HERA, Tevatron and fixed target data: MSHT20 PDFs}",
    eprint = "2012.04684",
    archivePrefix = "arXiv",
    primaryClass = "hep-ph",
    reportNumber = "IPPP/20/58",
    doi = "10.1140/epjc/s10052-021-09057-0",
    journal = "Eur. Phys. J. C",
    volume = "81",
    number = "4",
    pages = "341",
    year = "2021"
}

@article{Ball:2021leu,
    author = "Ball, Richard D. and others",
    title = "{The path to proton structure at 1\% accuracy}",
    eprint = "2109.02653",
    archivePrefix = "arXiv",
    primaryClass = "hep-ph",
    reportNumber = "Edinburgh 2021/12, Nikhef-2021-013, TIF-UNIMI-2021-11",
    doi = "10.1140/epjc/s10052-022-10328-7",
    journal = "Eur. Phys. J. C",
    volume = "82",
    number = "5",
    pages = "428",
    year = "2022"
}

@article{Alekhin:2017kpj,
    author = {Alekhin, S. and Bl\"umlein, J. and Moch, S. and Placakyte, R.},
    title = "{Parton distribution functions, $\alpha_s$, and heavy-quark masses for LHC Run II}",
    eprint = "1701.05838",
    archivePrefix = "arXiv",
    primaryClass = "hep-ph",
    reportNumber = "DESY-16-179, DO-TH-16-13",
    doi = "10.1103/PhysRevD.96.014011",
    journal = "Phys. Rev. D",
    volume = "96",
    number = "1",
    pages = "014011",
    year = "2017"
}

@article{LHCb:2018usb,
    author = "{LHCb Collaboration}",
    title = "{Measurement of forward top pair production in the dilepton channel in $pp$ collisions at $\sqrt{s}=13$ TeV}",
    eprint = "1803.05188",
    archivePrefix = "arXiv",
    primaryClass = "hep-ex",
    reportNumber = "LHCb-PAPER-2017-050, CERN-EP-2018-022, LHCB-PAPER-2017-050",
    doi = "10.1007/JHEP08(2018)174",
    journal = "JHEP",
    volume = "08",
    pages = "174",
    year = "2018"
}

@Booklet{PUB-LUMI-2023-01,
    author         = "{ATLAS Collaboration}",
    title          = "{Preliminary analysis of the luminosity calibration of the ATLAS 13.6 TeV data recorded in 2022}",
    howpublished   = "{ATL-DAPR-PUB-2023-001}",
    url            = "https://cds.cern.ch/record/2853525",
    year           = "2023",
}

@article{CMS:2023qyl,
    author = "{CMS Collaboration}",
    title = "{First measurement of the top quark pair production cross section in proton-proton collisions at $\sqrt{s}$ = 13.6 TeV}",
    eprint = "2303.10680",
    archivePrefix = "arXiv",
    primaryClass = "hep-ex",
    reportNumber = "CMS-TOP-22-012, CERN-EP-2023-021",
    month = "3",
    year = "2023"
}

@article{Bellm:2019zci,
    author = "Bellm, Johannes and others",
    title = "{Herwig 7.2 release note}",
    eprint = "1912.06509",
    archivePrefix = "arXiv",
    primaryClass = "hep-ph",
    reportNumber = "MAN/HEP/2019/011, CERN-TH-2019-213, IFJPAN-IV-2019-18, HERWIG-2019-02, UWTHPH-19-36, KA-TP-24-2019, LU-TP 19-57, MCnet-19-28, IPPP/19/91",
    doi = "10.1140/epjc/s10052-020-8011-x",
    journal = "Eur. Phys. J. C",
    volume = "80",
    number = "5",
    pages = "452",
    year = "2020"
    }

@article{LHCb:2015nta,
    author = "{LHCb Collaboration}",
    title = "{First observation of top quark production in the forward region}",
    eprint = "1506.00903",
    archivePrefix = "arXiv",
    primaryClass = "hep-ex",
    reportNumber = "LHCB-PAPER-2015-022, CERN-PH-EP-2015-132",
    doi = "10.1103/PhysRevLett.115.112001",
    journal = "Phys. Rev. Lett.",
    volume = "115",
    number = "11",
    pages = "112001",
    year = "2015"
}

@article{Catani:2009sm,
    author = "Catani, Stefano and Cieri, Leandro and Ferrera, Giancarlo and de Florian, Daniel and Grazzini, Massimiliano",
    title = "{Vector Boson Production at Hadron Colliders: A Fully Exclusive QCD Calculation at Next-to-Next-to-Leading Order}",
    eprint = "0903.2120",
    archivePrefix = "arXiv",
    primaryClass = "hep-ph",
    doi = "10.1103/PhysRevLett.103.082001",
    journal = "Phys. Rev. Lett.",
    volume = "103",
    pages = "082001",
    year = "2009"
}

@article{Li:2012wna,
    author = "Li, Ye and Petriello, Frank",
    title = "{Combining QCD and electroweak corrections to dilepton production in the framework of the FEWZ simulation code}",
    eprint = "1208.5967",
    archivePrefix = "arXiv",
    primaryClass = "hep-ph",
    reportNumber = "ANL-HEP-PR-12-68",
    doi = "10.1103/PhysRevD.86.094034",
    journal = "Phys. Rev. D",
    volume = "86",
    pages = "094034",
    year = "2012"
}
 
\clearpage
 
% ATLAS Collaboration author list
% Reference date of TOPQ-2023-21 is 2023-07-28
% Author list last updated on date 13-DEC-23
% Data extracted on 13-Dec-2023 for paper reference TOPQ-2023-21
% at 11:47pm
 
\begin{flushleft}
\hypersetup{urlcolor=black}
{\Large The ATLAS Collaboration}

\bigskip

\AtlasOrcid[0000-0002-6665-4934]{G.~Aad}$^\textrm{\scriptsize 102}$,
\AtlasOrcid[0000-0002-5888-2734]{B.~Abbott}$^\textrm{\scriptsize 120}$,
\AtlasOrcid[0000-0002-1002-1652]{K.~Abeling}$^\textrm{\scriptsize 55}$,
\AtlasOrcid[0000-0001-5763-2760]{N.J.~Abicht}$^\textrm{\scriptsize 49}$,
\AtlasOrcid[0000-0002-8496-9294]{S.H.~Abidi}$^\textrm{\scriptsize 29}$,
\AtlasOrcid[0000-0002-9987-2292]{A.~Aboulhorma}$^\textrm{\scriptsize 35e}$,
\AtlasOrcid[0000-0001-5329-6640]{H.~Abramowicz}$^\textrm{\scriptsize 151}$,
\AtlasOrcid[0000-0002-1599-2896]{H.~Abreu}$^\textrm{\scriptsize 150}$,
\AtlasOrcid[0000-0003-0403-3697]{Y.~Abulaiti}$^\textrm{\scriptsize 117}$,
\AtlasOrcid[0000-0002-8588-9157]{B.S.~Acharya}$^\textrm{\scriptsize 69a,69b,m}$
% ...
\bigskip
\\

% ...
$^{29}$Physics Department, Brookhaven National Laboratory, Upton NY; United States of America.\\
% ...
$^{35}$$^{(a)}$Facult\'e des Sciences Ain Chock, R\'eseau Universitaire de Physique des Hautes Energies - Universit\'e Hassan II, Casablanca;$^{(b)}$Facult\'{e} des Sciences, Universit\'{e} Ibn-Tofail, K\'{e}nitra;$^{(c)}$Facult\'e des Sciences Semlalia, Universit\'e Cadi Ayyad, LPHEA-Marrakech;$^{(d)}$LPMR, Facult\'e des Sciences, Universit\'e Mohamed Premier, Oujda;$^{(e)}$Facult\'e des sciences, Universit\'e Mohammed V, Rabat;$^{(f)}$Institute of Applied Physics, Mohammed VI Polytechnic University, Ben Guerir; Morocco.\\
% ...
$^{49}$Fakult\"{a}t Physik , Technische Universit{\"a}t Dortmund, Dortmund; Germany.\\
% ...
$^{55}$II. Physikalisches Institut, Georg-August-Universit\"{a}t G\"ottingen, G\"ottingen; Germany.\\
% ...
$^{69}$$^{(a)}$INFN Gruppo Collegato di Udine, Sezione di Trieste, Udine;$^{(b)}$ICTP, Trieste;$^{(c)}$Dipartimento Politecnico di Ingegneria e Architettura, Universit\`a di Udine, Udine; Italy.\\
% ...
$^{102}$CPPM, Aix-Marseille Universit\'e, CNRS/IN2P3, Marseille; France.\\
% ...
$^{117}$Department of Physics, New York University, New York NY; United States of America.\\
% ...
$^{120}$Homer L. Dodge Department of Physics and Astronomy, University of Oklahoma, Norman OK; United States of America.\\
% ...
$^{150}$Department of Physics, Technion, Israel Institute of Technology, Haifa; Israel.\\
$^{151}$Raymond and Beverly Sackler School of Physics and Astronomy, Tel Aviv University, Tel Aviv; Israel.\\
% ...
$^{a}$ Also Affiliated with an institute covered by a cooperation agreement with CERN.\\
$^{b}$ Also at An-Najah National University, Nablus; Palestine.\\
$^{c}$ Also at Borough of Manhattan Community College, City University of New York, New York NY; United States of America.\\
$^{d}$ Also at Center for High Energy Physics, Peking University; China.\\
$^{e}$ Also at Center for Interdisciplinary Research and Innovation (CIRI-AUTH), Thessaloniki; Greece.\\
$^{f}$ Also at Centro Studi e Ricerche Enrico Fermi; Italy.\\
% ...
$^{m}$ Also at Department of Physics, King's College London, London; United Kingdom.\\
% ...

\end{flushleft}

% Created with Glance <Atlas.Glance@cern.ch>

\end{document}